\documentclass[aip,reprint]{revtex4-1}
	
\usepackage{epsfig}
\usepackage{subfigure} 
\usepackage{amssymb,amsmath}
\usepackage{overpic}
\usepackage{epstopdf}
\usepackage[pdftex,bookmarks,colorlinks]{hyperref}

\newcommand{\be}{\begin{equation}}
\newcommand{\ee}{\end{equation}}
\newcommand{\bea}{\begin{eqnarray}}
\newcommand{\eea}{\end{eqnarray}}

\draft 

\begin{document}


\title{Perspective: Geometrically-Frustrated Assemblies} 



\author{Gregory M. Grason}
\email[]{grason@mail.pse.umass.edu}
\affiliation{Department of Polymer Science, University of Massachusetts, Amherst, Massachusetts 01003, USA}



\begin{abstract}
This perspective will overview an emerging paradigm for self-organized soft materials, {\it geometrically-frustrated assemblies}, where interactions between self-assembling elements (e.g. particles, macromolecules, proteins) favor local packing motifs that are incompatible with uniform global order in the assembly.  This classification applies to a broad range of material assemblies including self-twisting protein filament bundles, amyloid fibers, chiral smectics and membranes, particle-coated droplets, curved protein shells and phase-separated lipid vesicles.  In assemblies, geometric frustration leads to a host of anomalous structural and thermodynamic properties, including heterogeneous and internally-stressed equilibrium structures, self-limiting assembly and topological defects in the equilibrium assembly structures.  The purpose of this perspective is to 1) highlight the unifying principles and consequences of geometric frustration in soft matter assemblies;  2) to classify the known distinct ÒmodesÓ of frustration and review corresponding experimental examples; and 3) to describe outstanding questions not yet addressed about the unique properties and behaviors of this broad class of systems.

\end{abstract}


\maketitle 



%
%

%


\section{Introduction}
\label{sec:intro}

In canonical self-assembling systems, thermodynamics for allows for two states:  the dispersed state of constituent elements or a condensed state of unlimited dimensions.  This derives from the fact that equilibrium ordering in the condensed state is ``homogeneous" on some scale much larger than the building block or the periodic repeat of its crystalline packing. At large scales, thermodynamics in these systems reduces to a negative bulk free energy gain for assembly, $\sim - L^D$ (where $D$ and $L$ are respective dimensionality and size) that cannot be restrained at large $L$ by the positive surface energy cost at the boundary, $\sim  L^{D-1}$.   The focus of this perspective is on a class of self-assembling systems that break this fundamental paradigm, {\it geometrically-frustrated assemblies} (GFAs).  

\begin{figure}
 \epsfig{file=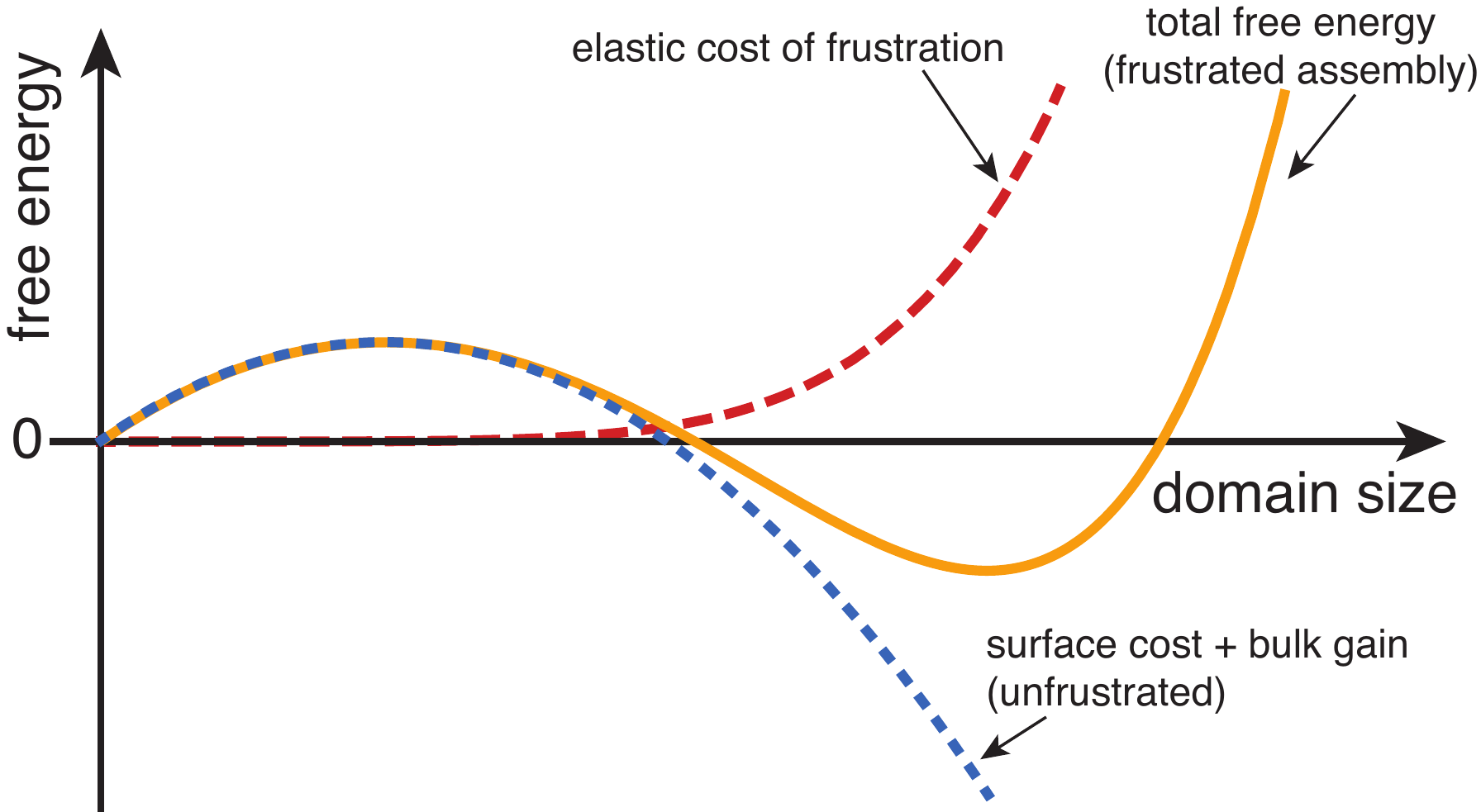, width=2.8in}%
 \caption{Schematic of size-dependent thermodynamics of GFAs.  The dotted blue line shows that free energy an unfrustrated assembly, while the dashed red line shows the additional elastic cost size-dependent frustration in GFAs. }
 \label{fig: FvsL}%
 \end{figure}

Geometric frustration, or the incompatibility of local interactions with global geometric constraints~\cite{sadoc_mosseri, kleman1987}, is most often associated with low-temperature ordering of spin systems in solid state, with the canonical example being the frustration of anti-ferromagnetic ordering on triangular lattices~\cite{binder_young1986}.  In the context of assemblies, which lack {\it a priori} lattice order, geometric frustration implies the inability to propagate a preferred pattern of local ordering globally.  Often this takes the form of assembled  ``building blocks" whose interactions and structure prefer two (or more) mutually incompatible patterns of order.  In rigid or brittle systems, such an incompatibility may simply prevent assembly altogether.  In soft systems the weaker non-covalent forces that hold together fairly large and flexible building blocks (proteins, macromolecules, colloids) can tolerate some measure of local misfit, allowing imperfect order to extend over at least some finite range.    As a consequence of the scale-dependent misfit in these assemblies, the formation of a domain of size $L$ generically comes with an additional thermodynamic cost, $\sim f_{frust} (L) L^D$, where $ f_{frust} (L)$ describes the free energy density cost of frustration (see Fig.~\ref{fig: FvsL}).  As $f_{frust}$ is an increasing function of $L$ (over some size range), the cost of frustration can in many situations dominate the bulk-energy gain of assembly for large $L$, and lead to structural and thermodynamic behaviors that are wholly distinct from canonical, unfrustrated assemblies.  Perhaps most significant among these is the equilibrium assembly of identical elements into structures of well-defined, {\it self-limiting} dimensions, which can be much larger than the building block size, yet finite (i.e. not bulk condensates).

This classification applies to a broad class of self-assembling materials systems, including self-twisting assemblies of chiral filaments (e.g. filamentous proteins, amyloids), chiral smectics and bilayer membranes, particle-coated droplets, curved protein shells (e.g. viral capsids) and phase-separated lipid vesicles. To date the implications of geometric frustration have only been studied in context of disperate, apparently unrelated material systems, and as a consequence, the ``anomalous"  behaviors of GFAs relative to canonical assembly systems, and their potential application to bottom-up routes to materials structure, remain largely unrecognized .  The broad goal of this article is to present a unifying perspective on GFAs that describes the common underlying principles and outcomes of geometric frustration, surveys the known distinct geometric mechanisms of frustration in assemblies and highlights current challenges to understanding and manipulating their properties.  In Sec. ~\ref{sec: example} I introduce a model of a particular GFA, 2D crystals on spherical surfaces, in order to illustrate the basic principles and outcomes of  geometric frustration in assemblies.  This example will be used to illustrate the anomalous behaviors of GFAs, including scale-dependent intrinsic stresses, self-limiting assembly, topological defects in equilibrium and shape-selection of anisotropic domains.  In Sec.~\ref{sec: modes} I will survey the known  geometric mechanisms, or frustration ``modes", of GFA and review the specific assembly systems associated with each mode.  In~\ref{sec: open}, I highlight several open challenges to understanding the emergent behaviors of GFAs and opportunities for engineering geometric frustration as a ``design principle" for synthetic self-assembling materials.

\section{Principles of geometric frustration in assembly: an illustrating example}
\label{sec: example} 

To illustrate the principles and outcomes of geometric frustration on assembly, we consider a particular example of an assembly with preferred 2D crystalline order and spherical surface shape.  As reviewed in the next section, variations of this model have been applied to crystalline assemblies of particles confined to spherical interfaces (e.g. coated-droplets)~\cite{bowick_nelson_travesset, bausch2003}, or instead, to membrane assemblies of subunits that prefer spherical shape (e.g. capsids~\cite{zandi2004, lidmar2003} or clathrin cages~\cite{mahsl1998, kohyama2003}.  For the present purposes, the formation of the 2D crystal with positive Gaussian curvature provides the simplest model to illustrate the many complex structural responses of an assembly to geometric frustration.

Modeling the structure and thermodynamics of GFAs poses particular challenges, stemming largely from the fact that the organization of subunits varies from place to place within the assembly.  In other words, geometrically frustrated structures run counter to the {\it principle of maximum symmetry}, which posits that arrangements with higher symmetry, like periodic lattices, are more likely to be low potential-energy states~\cite{wales}.  Frustration promotes the formation of  often subtle gradients in the local order~\cite{sadoc_mosseri}.  Therefore, assessing the optimal free energy arrangement of an GFA requires modeling of inhomogenous arrangements of subunits which can vary on length scales much larger than a single subunit size.  Moreover, under certain conditions frustration promotes the formation of topological defects in the packing, which ultimately results in GFAs (even for fixed size) having fairly rough energy landscapes populated by structurally-distinct but nearly-degenerate ground states.  For these reasons, exhaustively probing the thermodynamics of GFAs is difficult to achieve via traditional numerical sampling approaches, such as Monte Carlo or Brownian Dynamics simulation applied to discrete subunit assembly models.

The use of continuum theories, which describe the structure and thermodynamics over length scales much larger than the subunit size, are particularly valuable for predictions of GFA behavior.  As applied to GFAs, the essential assumptions of a continuum theory are two-fold.  First, in the absence of frustration, ground-state ordering is described by a uniform order parameter of appropriate symmetry to describe the nature of order (e.g. orientational, positional).   Second, the for sufficiently weak frustration, the cost of geometrically-imposed gradients of the order parameter can be accounted for by a generalized elasticity theory~\cite{chaikin_lubensky}.  This description relies on the existence of low-energy deformation modes of the order whose cost vanish with the deformation wavevector.  Such modes, known as Goldstone modes, are guaranteed to exist for any type of ordered assembly since local arrangements requires the breaking of some combination of continuous (rotational and translational) symmetries~\cite{chaikin_lubensky}.  The presence of these long-wavelength ``elastic" modes plays a key role in the size-dependent structure and thermodynamics of GFAs.

\subsection{Continuum model of spherically-curved, 2D crystals}

\begin{figure}
 \epsfig{file=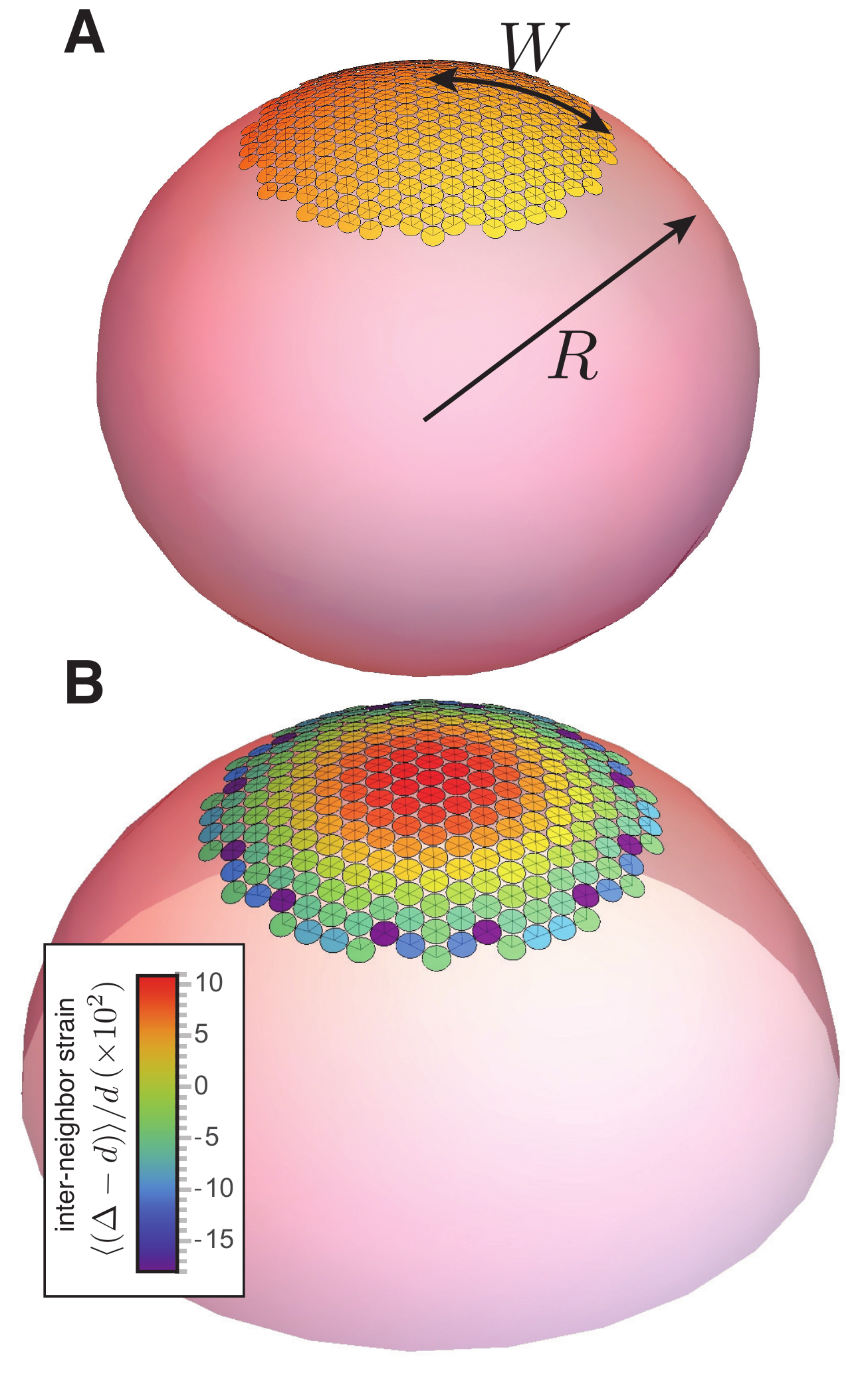, width=2.5in}%
 \caption{In (A), a 2D crystalline cap assembly with spherical curvature.  A map of the inter-element strain distribution is shown in (B) for simulated bead-spring cap for $2W=R=16 d$.  Colors indicate mean neighbor spacing $\langle \Delta\rangle$ relative to strain-free spacing $d$.  }%
 \label{fig: spheremodel}
 \end{figure}

As a prototype GFA, we consider a 2D crystalline domain of finite lateral size and a spherically-curved shape of radius, $R$ (see Fig. \ref{fig: spheremodel}A).  For simplicity, we will assume $R$ is constant throughout the assembly, as if imposed by a spherical substrate or instead by a uniformly tapered shape of assembly units.  Here, and elsewhere in the this perspective we assume the grand canonical ensemble, where subunits self-assemble at fixed chemical potential difference $\Delta \mu$ with a bath of dissociated unimers~\footnote{Assuming a cohesive energy $\epsilon$ per neighbor bond, $\Delta \mu=-3 \epsilon -k_B T \ln c_0$, where $c_0$ is the free unimer concentration}.   For this model, we consider the free-energy dependence of a cluster of assembled particles on the size and shape of the domain, possessing three contributions,
\begin{equation}
\label{eq: fsphere}
F=-\Delta \mu \rho_0 A + \Lambda P + F_{frust},
\end{equation}
where the first term represents the (2D) bulk gain of assembly of area $A$ with areal density $\rho_0$ and the second term represents the penalty of fewer cohesive bonds along a perimeter of length $P$, described by line tension $\Lambda$.  These first two terms are the standard ingredients of a  classical 2D nucleation model.  The third term $F_{frust}$ described the size- and shape-dependent cost introduced by frustration.  Here, we consider a simplified continuum model based on the F\"oppl-v\'an K\'arman theory of curved 2D crystalline membranes~\cite{seung1988, bowick_giomi2009} ,
\begin{equation}
\label{eq: frust}
F_{frust}=\frac{1}{2} \int dA ~\Big[ \sigma_{ij} u_{ij} +C (K_g -K_0)^2 \Big] ,
\end{equation}
where $\sigma_{ij}$ and $u_{ij}$ are the in-plane elastic stress and strain associated with deformations from uniform crystalline order.  The second term in eq. (\ref{eq: frust}) introduces a phenomenological cost for change in the Gaussian curvature~\footnote{The Gaussian curvature of a surface is the product of the two principle curvatures of the surface, and equal to $+R^{-2}$ for sphere~\cite{kamien2002}.}, $K_g=+R^{-2}$, of the assembly away from its preferred value, $K_0 = +R_0^{-2}$.  Here, we take $C$ to be an effective stiffness associated with this shape change, which may be a proxy for the cost of deforming tapered building block shape (as in the case of capsid-like assemblies~\cite{morozov2010}) or instead the cost of flattening a deformable curved substrate (such as the interface of a liquid droplet), though it should be emphasized that this model for shape change is not intended as a realistic description for either model.  Instead, we introduce this  phenomenological model to illustrate the differences between ``hard" and ``soft" frustration.  In the former case of ``hard" frustration (for $C \to \infty$), the imposed Gaussian curvature is rigid and the cost of frustration is unavoidable at all scales, whereas for ``soft" frustration (finite $C$) a finite cost for ``defrustrating" the assembly allows for adjusting the balance of the the size-dependent costs introduced by frustrated shape.

\subsection{Anomalous behaviors}

Here, we survey the anomalous behaviors of GFAs, illustrated by way of the continuum model of spherical crystalline domains introduced above.  These behaviors are ``anomalous" in the sense that they are not exhibited by equilibrium assembly canonical systems, composed of identical elements interacting through finite-range cohesive forces (i.e. finite on the scale of nearest-neighbor distances in the assembly).  For comparison, consider the planar, unfrustrated ($K_g=K_0=0$) case of the model above.  In this case, assemblies are strain-free and $F_{frust}=0$.  Ignoring the effect of possible faceting at the crystal boundary, we may assume 2D crystallites to have circular shape to minimize the cost of boundary.  The probability of forming a roughly circular domain of radius $W$ is simply $P(W) \propto \exp\{ - F(W)/k_B T\}$~\cite{israelachvili}.  When $\Delta \mu <0$ the bulk free energy favors the dispersed state, equilibrium domain sizes are distributed according the to tail of a Gaussian distribution, with the largest population from unimers (i.e. $P(W)$ is peaked at $W\to 0$).  Under conditions favoring condensation, $\Delta \mu >0$, $F(W)$ decreases without bound as $W \to \infty$, indicating that equilibrium behavior favors a single bulk aggregate.  This example, and other cases of canonical assemblies, only admit two equilibrium phases:  a dispersed state for $\Delta \mu <0$ dominated by unimers and a distribution of aggregates that decays exponentially with size; and a condensed state for $\Delta \mu>0$ dominated by a single, macroscopic aggregate.  We contrast this canonical case with the behavior of GFA below.

\subsubsection{Inhomogeneous local packing and intrinsic stress}

Periodic positional 2D order is generically incompatible with surfaces of with Gaussian curvature, like spheres~\cite{rubinstein_nelson1983, kleman1989, bowick_giomi2009}.  We can understand this from the fact that straight lines (geodesics) on such surfaces are not parallel:  geodesics converge, on surfaces like globes, and diverge, on saddles~\cite{kamien2002}.  Picturing those ``straight curves" as the rows of a 2D lattice, it is intuitive to imagine that extending order on the surface generates transverse inter-element strains that grow with row length.  This geometrical effect is captured by the so-called compatibility equation that relates gradients of in-plane stress in an elastic membrane to its out-of-plane geometry~\cite{seung1988},
\begin{equation}
\label{eq: compat}
\frac{1}{2 Y} \nabla^2_\perp \Pi ({\bf x}) = K_g - s ({\bf x}) ,
\end{equation}
where $\Pi \equiv -\sigma_{ii}/2$ is the lateral elastic pressure in the assembly,  $Y$ is the 2D Youngs modulus self-assembled array and $s ({\bf x})$ is the areal density of disclinations (e.g. 5- or 7-fold defects in an hexagonal array).  The compatibility relation shows that Gaussian curvature of an assembly imposes gradients in the stress akin to a {\it continuous distribution of defects}~\cite{kleman1989}.  

Below we describe the effect of curvature to stabilize defects.  Here, we first consider the defect-free case of a circular domain (a ``cap" of lateral radius $W$ shown in Fig.~\ref{fig: spheremodel}) to illustrate the variation of local packing in the frustrated assembly. Taking $s ({\bf x}) =0$, eq. (\ref{eq: compat}) implies that spherical curvature ($K_g = + R^{-2}$) requires the assembly to become increasingly compressive with radial arc distance from the cap center, $r$, with $\Pi(r) = \Pi(0) + Y K_g r^2/2$ (see strain distribution in Fig.~\ref{fig: spheremodel}B).  Again, this compression is intuitive when considering areal strain in the assembly $u_{ii} = -(1-\nu) \Pi/2$, where $\nu$ is the 2D Poisson ratio of the hexagonal array .  A uniform-density hexagonal array has  $dN(r) \simeq 2 \pi r \rho_0 dr$ elements a distance $r$ from the center.  On a sphere, these are confined to an area $dA=\ell(r) dr$, where the length of a latitude an arc-distance $r$ from the from the pole is $\ell(r) = 2 \pi R \sin(r/R) \simeq 2 \pi r \big[1-r^2/(6R^2)\big]$.  Because the amount of room available on the sphere increases more slowly with $r$ than the number of sites in the lattice~\cite{nelson_spaepen1989}, the density of elements in the cap must grow with distance from the center:  $\rho(r) =dN(r)/dA \simeq \rho_0 \big[1+r^2/(6R^2)\big]$ for $r \ll R$.

\begin{figure*}
 \epsfig{file=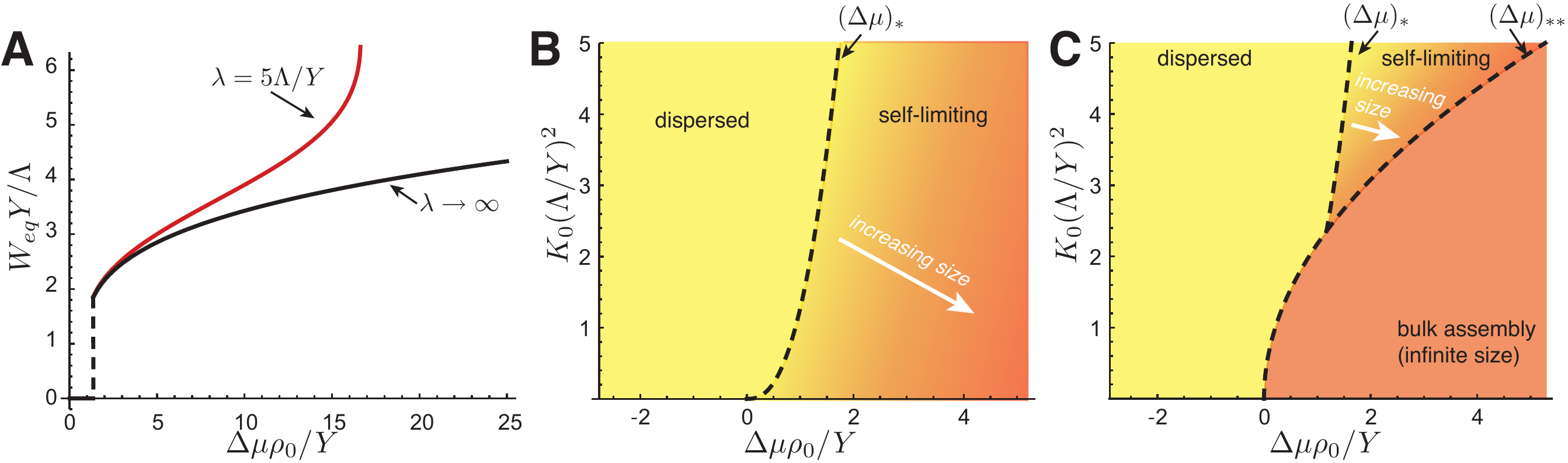, width=7in}%
 \caption{In (A), plots of equilibrium cap size versus bulk chemical potential for $R=\sqrt{3} \Lambda/Y$; black and red curves show ``hard" and ``soft" frustration, respectively.  Cap assembly phase diagrams are shown for ``hard" frustration in (B) and for ``soft" frustration (with $\lambda=3 \Lambda/Y$) in (C). In (B) and (C), $(\Delta \mu)_*$ labels the boundary between the dispered and self-limiting phase and $(\Delta \mu)_{**}$ is the transition to the bulk assembly phase. }%
 \label{fig: sphtherm}
 \end{figure*}

The condition of non-zero Gaussian curvature requires the local packing to vary throughout the assembly, even in a defect-free assembly.  Despite the generic preference for uniform order, inter-element packing at the edge of cap is more dense than the interior by a factor that grows as $(W/R)^2$.  Hence, the local arrangements, as characterized by deformations of the preferred hexagonal packing, not only vary throughout the assembly, but also from one assembly to another, according to the size of the assembly.  As illustrated below, the size-dependence of these intrinsic stresses in GFAs has critical implications for assembly thermodynamics.

\subsubsection{Self-limiting equilibrium assembly}

Here, we demonstrate that geometric frustration leads to self-limiting equilibrium assembly.  That is, equilibrium assemblies have a preferred size, which may be much larger than the subunit size, yet finite. To clarify, this behavior is distinct from other classes of assembly forming equilibrium aggregates of self-limiting dimension.  On one hand, GFA requires no long-range interactions, unlike so-called {\it short-range attraction, long-range repulsion} models in which equilibrium dimensions of a self-limiting cluster cannot substantially exceed the maximum range of repulsive interactions~\cite{groenewold2001}, such as the electrostatic screening length.  In GFA, interactions may be strictly local (i.e. nearest neighbor), yet as demonstrated above, lead to {\it stresses} in the assembly that are long-range compared to the subunit size.  Further, self-limitation in GFA is distinct from the well-known case of amphiphile assembly, where the ``head-tail" polarity of molecular unit leads to the equilibrium formation aggregates with some preferred dimensions.  In this latter case, the finite dimensions of the aggregates (e.g. radius of micelles or bilayer thickness) is limited by the size of the molecular unit (e.g. surfactant~\cite{israelachvili} or amphiphilic polymer~\cite{blanazs2009}).  In comparison, due to the long-range stresses in GFA, their self-limiting sizes can be {\it arbitrarily much larger than constituent element dimensions}, in principle, by one or more orders of magnitude.

For the case of spherical crystalline caps, this self-limitation manifests itself as an equilibrium, finite lateral dimension of the crystal, determine by the relative bulk free energy gain for assembly and cost of geometric frustration~\cite{schneider2005}.  Provided the equilibrium dimension is sufficiently small, we may assume that the crystal is defect-free and maintains an approximately circular shape.  In this case, the F\"oppl-van K\'arman theory described above predicts 
\begin{equation}
\frac{F_{frust}(W, K_g)}{ \pi W^2} = \frac{Y}{384} K_g^2 W^4 + \frac{C}{2} (K_g-K_0)^2 ,
\end{equation}
where the first term describes of the size-dependent elastic cost imposed by geometric frustration.  By minimizing with respect to curvature, it is straightforward to see that  domains of larger size flatten according to 
\begin{equation}
\label{eq: kgeq}
K_g^{eq} (W) = \frac{ K_0}{1+(W/\lambda)^4} ,
\end{equation}
where $\lambda \equiv \big( \frac{192 C} {Y} \big)^{1/4}$ is a length scale that parameterizes the cost of defrustrating the assembly through shape change.  For a circular cap, the free-energy then has be the following size dependent form
\begin{equation}
F(W) = - \pi \Delta \mu \rho_0 W^2 + 2\pi \Lambda W + \frac{ \pi Y K_0^2 W^6}{384\big[1+(W/\lambda)^4\big]} .
\end{equation}
Note the effect of the geometric frustration gives rise to a free-energy cost for domain formation, the final term above, that is monotonically increasing with $W$.  

Considering first the case of ``hard" frustration where $\lambda \to \infty$ and $K_g \to K_0$, it is straightforward to show that the size dependent cost affects the onset of 2D aggregation as indicated by the condition of $F(W)<0$ for some value of $W>0$.  While in the absence of frustration the onset of aggregation takes place at $\Delta \mu =0$, for finite curvature the transition from unimer-dominated to aggregate dominated occurs for $(\Delta \mu)_* = 6/5 Y^{1/5} \Lambda^{4/5} (K_0/2)^{2/5}$, showing the effect of increased frustration to obstruct assembly altogether (see Fig.~\ref{fig: sphtherm}B).  For much larger $\Delta \mu$, far in excess of $(\Delta \mu)_*$, minimum of $F(W)$ is always negative and occurs for a cap radius,
\begin{equation}
W_{eq} \simeq \Big ( \frac{ 128 \Delta \mu \rho_0}{Y} \Big)^{1/4} R_0;  \ \ {\rm for \ } C \to \infty \ {\rm and} \ \Delta \mu\gg (\Delta \mu)_* ,
\end{equation}
where we may neglect the effect of line-tension because $W_{eq} \gg \Lambda / (\rho_0 \Delta \mu)$.  The presence of a minimum of $F(W)$ for a finite $W$ indicates a {\it distinct phase of self-limiting domains}, with a roughly-Gaussian size distribution peaked at $W=W_{eq}$.

For the case of ``soft frustration", when domain size is comparable to or larger than $\lambda$,  assembled caps tend to flatten according to eq. (\ref{eq: kgeq}).  Overall, this ability to defrustrate the assembly through shape changes leads to a softening of the size-dependent costs of assembly, and overall a larger size of self-limiting domains, as shown in Fig.~\ref{fig: sphtherm}A.  Near to the critical line for the onset of aggregation, this leads to only a slight depression of $(\Delta \mu)_* $ relative to the ``hard frustration" limit described above.  More significant, the ability to completely flatten in the $W/\lambda \to \infty$ limit, implies that the free energy of the domain will ultimately grow with the area, with a ``renormalized" bulk free-energy density $-\Delta \mu \rho_0+C K_0^2/2$ that includes the added cost of relaxing preferred curvature.  Therefore, for large curvature, $K_0 > K_{c} \approx 40 \sqrt{\Lambda/Y \lambda^5}$, finite-size assemblies only occur for intermediate chemical potentials, $(\Delta \mu)_* >\Delta \mu >(\Delta \mu)_{**}$, where the upper-limit is simply $(\Delta \mu)_{**} = \rho_0^{-1} C K_0^2/2$.  For weak frustration (or small curvature $K_0 < K_{c}$) the onset of finite-size domain formation is preempted by the stability of bulk (flattened) domains of unlimited size, and the range self-limiting assembly at intermediate chemical potential is absent.  Figs.~\ref{fig: sphtherm}B and C show the respective phase diagrams for the ``hard" and ``soft" frustration models.

\subsubsection{Defects in the ground-state order}

The limitation of equilibrium dimensions is just one of multiple structural responses to the size-dependent elastic costs generated by geometric frustration.  A second such response is the formation of topological defects in the ground-state order, a behavior that is also characteristic of bulk ordering in frustrated systems, such as the liquid crystal blue phases~\cite{kleman1987, wright_mermin1989}.  Topological defects represent singular, point-like or line-like disruptions in the order parameter that cannot be undone by smooth and continuous deformations~\cite{chaikin_lubensky, nelson}.  The model of 2D crystalline order described here allows for both orientational and positional topological defects, which are disclinations and dislocations, respectively.  As shown in eq. (\ref{eq: compat}) Gaussian curvature has a similar effect to a ``smeared out" distributions of disclinations, disrupting orientation through the convergence or divergence of lattice directions~\cite{kleman1989}.  Therefore, it is natural to consider that the cumulative disorienting effect of Gaussian curvature can, be compensated, or ``neutralized", to some extent by the disclinations of the appropriate sign~\cite{nelson_peliti1987, lubensky_prost1992}.  

\begin{figure}
 \epsfig{file=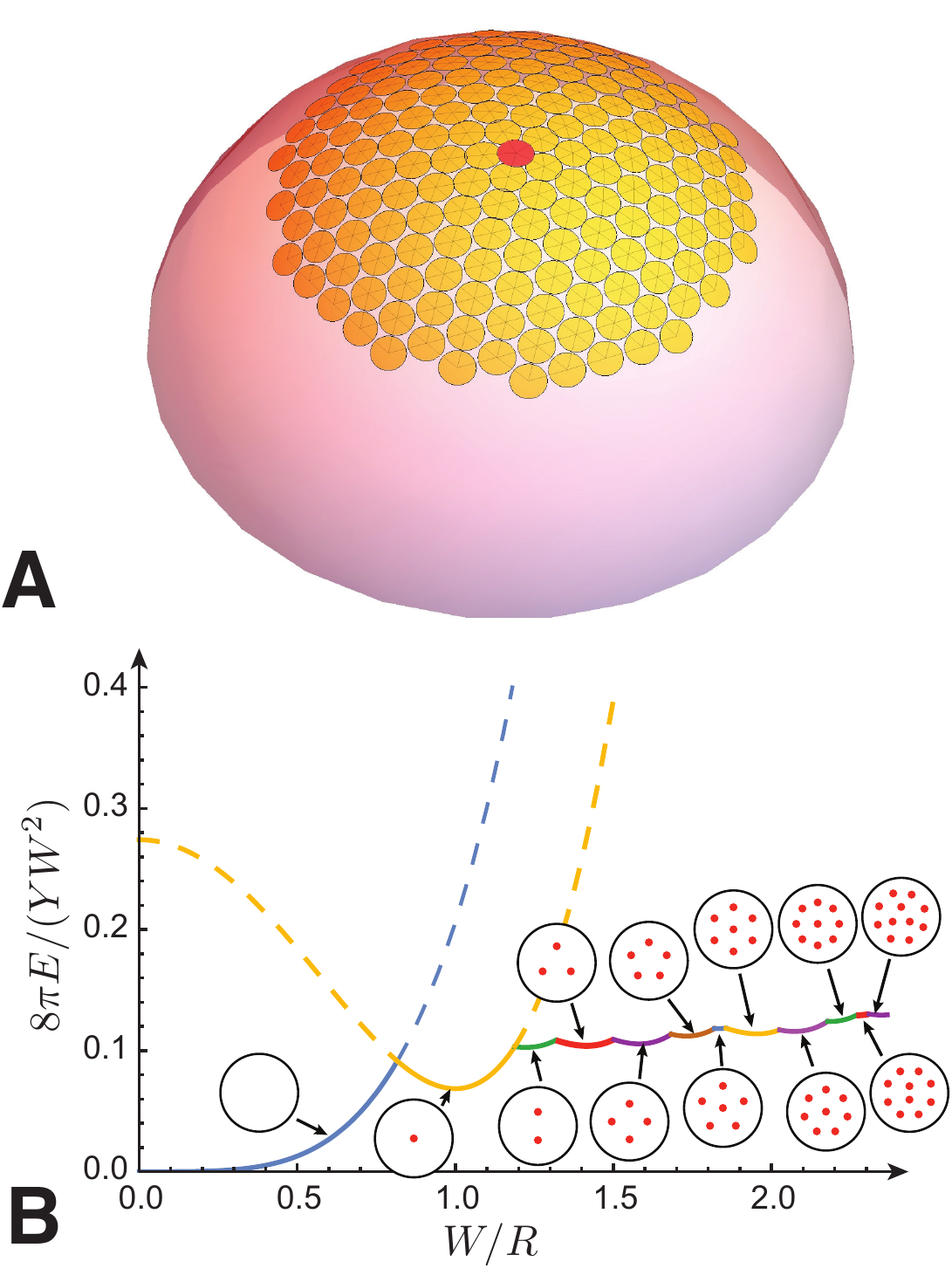, width=2.75in}%
 \caption{(A) shows a cap with a central 5-fold disclination (red).  The elastic energy density of a cap is shown as function of increasing curvature is shown in (B).  Piecewise-continuous solid portions indicate favorable configurations with integer numbers of disclinations (results adapted from ref.~\cite{grason2012}). Dashed portions show metastable branches for defect-free and single 5-fold disclination caps.}%
 \label{fig: defect}
 \end{figure}

The stability of topological defects in curved crystals, and GFA more broadly, represents an even more profoundly heterogeneous  ``solution" to the problem imposed by geometric frustration than the smooth distortions of uniform packing describe above.  In a hexagonal lattice, 5-fold (7-fold) disclination defects correspond to the removing (adding) of a $60^\circ$ angular wedge to  crystal then closing, leading to 6-fold coordination everywhere except but for a 5-fold (7-fold) site at the defect core (see Fig.~\ref{fig: defect}A).  Disclinations are characterized by  a {\it topological charge}, $s$ that quantifies the angular rotation of lattice directions around loops that enclose the defects, with $s = \pm  \pi/3$ corresponding 5- and 7-fold defects, respectively.  From eq. (\ref{eq: compat}), we see that positive disclinations have the effect of partially relaxing the stresses generated by positive Gaussian curvature.  Heuristically, the removal of a material wedge implies that 5-fold defects generate tension in the crystal at far field, which tends to relax the geometric compression imposed by curvature.  

Considering a centered disclination in a crystalline cap, corresponding to $s({\bf x}) = s \delta ({\bf x})$, solving the compatibility equation and equations of mechanical equilibrium give the elastic energy of a ``defective" cap~\cite{azadi2016},
\begin{equation}
\frac{1}{2} \int dA ~ \sigma_{ij} u_{ij} = Y \pi W^2 \Big[ \frac{ K_g^2 W^4 }{384}  - \frac{s K_g W^2}{ 64 \pi}+ \frac{s^2}{32 \pi^2}\Big] .
\end{equation}
When $W \geq \sqrt{2/3}R$, a 5-fold disclination ($s = +\pi/3$) is favorable over the defect-free cap ($s=0$) in terms of the elastic energy~\footnote{Note we have not included an addition core energy, $E_c$, which accounts for corrections to the linear elastic cost of disclinations near the microscopic core~\cite{chaikin_lubensky} and will shift the threshold curvatures to for stable defects to higher values than predicted by the elastic energy alone.  As one typically expects this energy to scale with the microscopic lattice dimension $a$ as $E_c \sim Y a^2$, the contribution of the core energy relative to the elastic energy is proportional to $(a/W)^2$ and may be neglected when  $W \gg a$.}. 

At higher curvatures, (or larger $W$) where we expect additional disclinations, the complex spectrum of multi-defect cap ground states may be predicted by assuming the cap maintains the circular shape and extending this F\"oppl-van K\'arman calculation to multiple disclinations~\cite{grason2012} (see Fig.~\ref{fig: defect}B).  The stability of exactly 12 excess 5-fold disclinations in {\it closed spherical shells} is well-known in the context of the famous Thompson problem~\cite{weaire, bowick_giomi2009, altschuler1997}. Fig.~\ref{fig: defect}B shows that progression from the defect-free order in flat caps to multi-defect ground states with increasing spherical surface coverage leads to a highly non-linear dependence of $F_{frust}$ on assembly dimensions $W$.  The transitions between states of different integer numbers of disclinations lead to cusp-like features in the energy landscape, where the resistance to increasing the cap size (i.e. the slope of the elastic energy) drops abruptly due to the relaxation provided by the added defect.  Revisiting the self-limiting assembly behavior, the presence of favorable topological defects which partially relax the cost of geometric frustration have the effect allowing crystalline domains, and GFA more broadly, to grow to larger dimensions than they would in the absence of defects.  Furthermore, the cuspy nature of the energy dependence on domain size will have the effect of introducing discontinuities in the dependence of equilibrium dimensions on chemical potential for assembly corresponding to discretized ranges of stable domain size possessing different defect numbers.  

Because crystals possess both discrete translational and rotational symmetries, they also support positional defects, or {\it dislocations}.  Edge dislocations correspond to the addition or removal of a partial row of lattice positions, and in an hexagonal crystal, dislocation cores can be decomposed into a ``topologically neutral" 5-7 pair of disclinations~\cite{chaikin_lubensky}.  While ``charged" disclinations are most commonly studied as mechanisms of frustration relaxation in curved crystals (and they are topologically necessary in closed shells), dislocations in appropriate orientation and location in the crystal are also effective in relaxing geometric stresses in curved crystals~\cite{vitelli2006, irvine2010, azadi2014}.  Which defect type is preferred in a curved crystal is sensitive to the lattice spacing~\cite{azadi2016}.  The removal of a partial row of width $a$ and length $~W$ relaxes the geometric compression of the crystal by roughly $Y a W^3/R^2$.  Comparing this to the far-field elastic cost of dislocations, which grow as $Y a^2 \ln (W/a)$~\cite{chaikin_lubensky}, dislocations become elastically favorable when curvature $(W/R)^{2}$ exceeds a critical value $\approx W/a \ln (W/a)$~\cite{azadi2012}.  When lattice spacings $a$ are comparable to (but less than) crystal dimension $W$, disclinations are more efficient means of elastic energy.   One the other hand, when crystal width far exceeds the lattice spacing, dislocation stability with increasing curvature preempts disclinations, and defect ground states possessing a number of dislocations that grows as $\sim W/a$, which self-organize into chains, ``scars" or ``pleats"~\cite{irvine2010, grason_davidovitch2013, azadi2014, wales2013}.  Similar arguments show that for $W/a \gg 1$ chains of dislocations also decorate ``charged" defect configurations, that is, those possessing an excess 5- or 7-fold disclination~\cite{bowick_nelson_travesset, perez-garrido1997, wales2009}. Understanding how theses more complex patterns of multi-dislocation ground states evolve with crystal size, how ``neutral" defect patterns compete with ``charged" defect patterns~\cite{azadi2016}, and further, how these defects shape the energy landscape of self-assembled curved crystals remains an area of active research.

\subsubsection{Anisotropic domain shape selection}

The previous sections illustrated the incorporation of singular defects as a structural response of the bulk of an assemblies to geometric frustration.  Here, we illustrate a distinct response associated with the shape of a free boundary of a self-assembling domain.  The above discussion of the crystalline cap considers the case of line tension $\Lambda$, large enough to maintain a roughly circular shape of the crystal edge.  When this line tension falls sufficiently, geometric frustration can give rise to domain shapes that break this isotropic symmetry, and introduce excess boundary.  In the case of crystals on spherical surfaces, this has been argued~\cite{schneider2005}, and shown experimentally~\cite{meng2014}, to lead to formation of ribbon-like crystalline domains that wrap the sphere along great circles.

\begin{figure}
 \epsfig{file=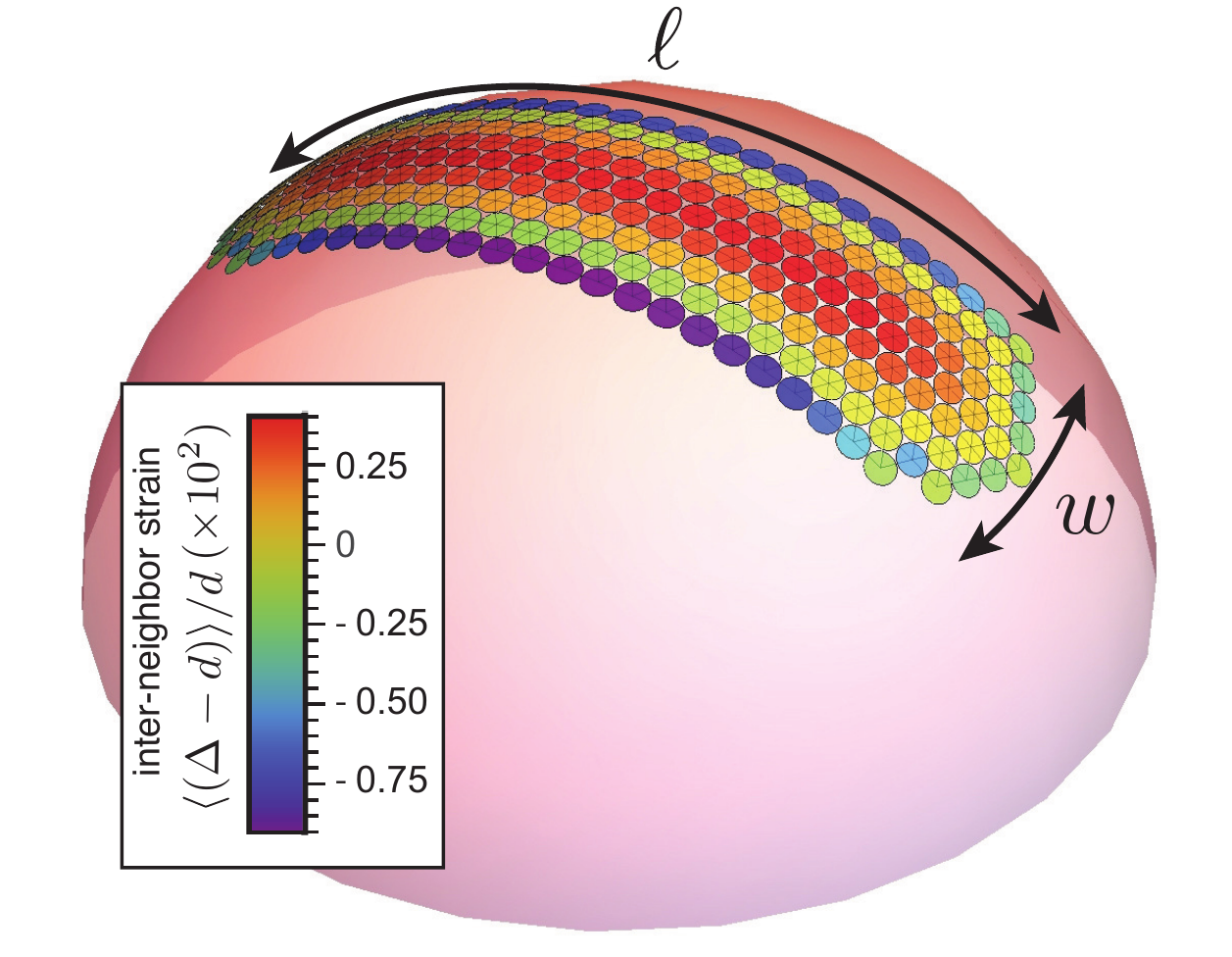, width=2.5in}%
 \caption{A map of the strain energy distribution for a ribbon-like 2D crystal from a simulated bead-spring model, shows a strain variation along the narrow direction.  The magnitude of strain is greatly reduced relative to the circular cap in Fig. 2, though of equal curvature and disc number. }%

\label{fig: strip}
 \end{figure}

Underlying this instability to low-symmetry domain shapes in GFAs is the effect of free boundaries to relax the stresses imposed by geometric frustration.  For 2D crystals on spheres this can be understood by considering pressure distribution in defect-free rectangular (belt-like) ribbon domain of length $\ell$ and transverse width $w$, such that $\ell \gg w$~\cite{schneider2005}.  Far away from the ends of the strip, the pressure depends only on the position along the (thinner) direction $y$ as $\Pi(y)\simeq \Pi (0) +Y K_g y^2/2$ (see Fig.~\ref{fig: strip}).  While the compatibility condition requires gradients (second derivatives) of pressure in proportion to $K_g$, this result shows that the pressure build-up is limited by the thinnest dimensions (i.e. $y=\pm w/2$) due to additional relaxation possible at the free boundary.  In this case, the cost of frustration becomes much more sensitive to $w$ than to $\ell$:  $F_{frust} \simeq Y \ell w^5 K_0^2 /1440$ (where we consider the ``hard frustration" limit for simplicity, $K_g=K_0$).  Considering a domain of fixed area $A=w \ell$ the optimal shape is determined by balance between this elastic cost and the line tension $\approx 2 \Lambda A/w$  (assuming $\ell \gg w$),  the minimal-energy aspect ratio of the ribbon becomes
\begin{equation}
\Big(\frac{ \ell}{w}\Big)_{min}\simeq 4 \pi \Big(\frac{Y R_0}{720 \Lambda} \Big)^{2/5} \Phi
\end{equation}
where $\Phi=w \ell/(4 \pi R^2)$ is the area fraction of the sphere covered by the domain.  The ratio $\Lambda/Y$ is a length scale, we may call it the {\it elasto-cohesive} length, that parameterizes the relative cost of fewer cohesive bonds at the edge of the domain to the elastic stretching of bonds in the interior of the domain.  In the limit where the cohesive interactions are relatively stiff (or brittle) $Y R_0/\Lambda \gg 1$ and the relatively small cost of excess perimeter allows domains to limit the cost of geometric frustration through highly anisotropic domain shape $(\ell/w)_{min} \gg1$.  Using $\Big(\frac{ \ell}{w}\Big)_{min} =1$ to estimate the transition between isotropic and anisotropic shapes, surface tension favors circular boundaries for domain dimensions less than roughly $\approx 2 (\Lambda/Y)^{1/5} R_0^{4/5}$, while domains spread laterally in response to frustration when the thinnest dimensions exceed this value.

The ability of free boundaries to circumvent the growth of frustration costs through anisotropic growth has important thermodynamic consequences for assembly at fixed chemical potential.  Neglecting the width-dependence of $P$ in the $\ell\gg w$ limit, the equilibrium width of the ribbon is $w_{eq}\simeq (288\Delta \mu \rho_0/Y)^{1/4}R_0$, indicating that the narrow dimension of the ribbon is retains roughly the same size as the circular cap diameter.  For this equilibrium width, the free energy of ribbon has the simple form
\begin{equation}
F(w_{eq},\ell) \simeq \Big[2\Lambda -\frac{4}{5} \mu \rho_0 w_{eq} \Big] \ell .
\end{equation}
Thus, when equilibrium width exceeds a critical value, the free energy per unit length of the ribbon is negative and $\ell$ grows unbounded~\cite{schneider2005, meng2014}, at least until the point of self-intersection ($\ell = 2 \pi R_0$).  Thus, the ability to reform the boundary of the domains of GFAs leads to equilibrium behavior that is intermediate to frustrated and unfrustrated assembly:  assemblies that are {\it self-limiting} in one or more of the dimensions, yet remain unlimited in others.

\section{Modes of geometrically frustrated assembly}
\label{sec: modes} 

The previous section reviewed a specific model of spherically-curved 2D crystals in order to overview the structural and thermodynamic responses to geometric frustration.  These behaviors -- gradients in long-range order, self-limiting dimensions, topological defects in the ground-state order and anisotropic domain selection -- are generic to GFAs, and have been studied in the context of a diverse range of self-assembly systems.  In this section, I review these distinct systems, both their experimental realizations and models, and classify them according the underlying geometric mechanism, or ``mode", of frustration.  Where possible, some attempt is made to summarize the key ingredients in the free energy that underlie the anomalous assembly behaviors in each correspond frustration mode, in a form corresponding to $F_{frust}$ introduced above for curved 2D crystals.  The following discussion makes exgensive reference to the various states of liquid-crystalline order realized in soft matter assemblies, which are intermediate to periodic crystals and isotropic liquids in terms of broken symmetries and response properties.  I refer readers who are unfamiliar with the basic taxonomy and classification of this ``liquid-crystal zoo" to an introductory text on the topic, such as ref. \cite{degennes_prost}, \cite{chaikin_lubensky} or ~\cite{kleman_lavrentovich}.

\subsection{Intrinsic curvature vs. long-range order in 2D assemblies}

The example illustrated above falls into the first category of GFA:  2D assemblies with non-zero intrinsic curvature.  Generically, A 2D surface is locally described by the principles curvatures, $\kappa_1$ and $\kappa_2$, measured in orthogonal directions~\cite{kamien2002}.  To clarify, a surface (a membrane or a substrate) is {\it intrinsically-curved} when is it has non-zero Gaussian curvature, $K_g = \kappa_1 \kappa_2 \neq 0$.  While {\it extrinsic curvature} of the surface, as described by, say, the mean curvature $H=(\kappa_1 + \kappa_2)/2$, may impose costs on assemblies due to shape-dependent strains on the assembly~\cite{mbanga2012}, the requirement of non-zero extrinsic curvature does not, itself, introduce geometric frustration.  For example, in a model where a membrane assembly has preferred mean curvature $H=H_0$, assemblies with $K_g=0$ (i.e. cylinders) achieve this geometry without introducing gradients in the long-range ordering of sub-units~\cite{napoli2012}.

\begin{figure*}
 \epsfig{file=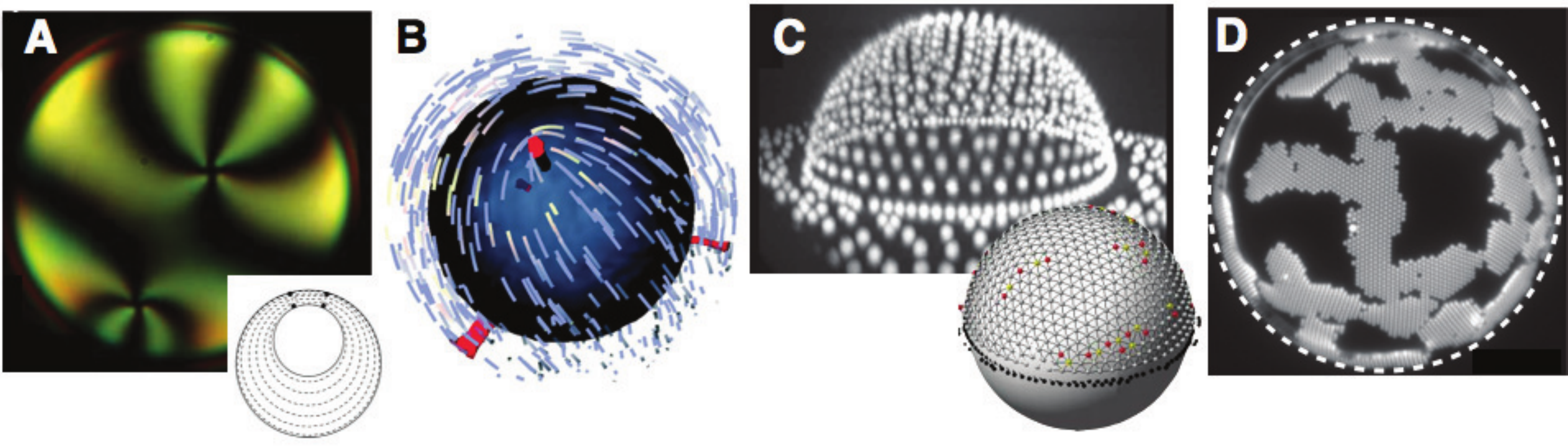, width=7in}%
 \caption{A polarized microscope image on a nematic shell within double emulsion drop (schematic inset) is shown in (A), adapted with permission from Phys. Rev. E {\bf 79}, 021707 (2009). Copyright American Physical Society (2009).  In (B), a simulation of a nematic shell layers, with disclinations shown in red, adapted with permission from Phys. Rev. Lett. {\bf 100}, 197802 (2008).  Copyright American Physical Society (2008).  In (C), a confocal microscope image of a partially wetting oil-water droplet coated by charged colloids, and corresponding defect reconstruction below, adapted with permission from Nature {\bf 468}, 947 (2010). Copyright Nature Publishing Group (2011).  In (D), a a confocal microscope image of branched crystal assembled at the interface of a spherical water-oil droplet from polystyrene particles with short-ranged cohesive interactions~\cite{meng2014}, where the dashed lines indicates the boundary of the 33 ${\rm \mu m}$ radius droplet (image courtesy of G. Meng and V. Manoharan). }%
 \label{fig: intrinsiccurv}
 \end{figure*}

\subsubsection{Orientational frustration}
Underlying the frustration of intrinsically-curved 2D assemblies is a classical result of differential geometry.  When $K_g \neq 0$ straight (geodesic) lines are not parallel, since they do not maintain constant spacing or relative angles~\cite{millman_parker}.  This is familiar from looking at the globe ($K_g= + R^{-2}$) on which the lines of longitude are geodesics, or great circles, that converge when traveling from the equator to the north pole where they intersect.  Conversely, geodesics diverge on $K_g <0$ surfaces, like saddles.  

For 2D assemblies that adopt intrinsically-curved shapes, this implies a frustration of {\it orientational order} between subunits.  Hence, assemblies with at least 2D {\it liquid crystalline} ordering are frustrated by curvature~\cite{nelson_peliti1987, nelson}.  Many models of assembly fall into this class.  Most directly related are models of nematic shells (see Fig.~\ref{fig: intrinsiccurv}A-B), which are prepared by confining a thin layer of nematic liquid crystal between two concentric liquid droplet interfaces, where mesogens have sufficiently strong tangential anchoring conditions with droplet surfaces~\cite{vitelli_nelson2006, skacej2008, lopez-leon2011, lopez-leon_nieves2011}.  A second class of models considers in-plane liquid-crystalline order as the result of membrane assembly~\cite{lubensky_prost1992, hirst2012}.  For example, for bilayer lipid membranes, which form curved vesicles under many conditions, in-plane anisotropy arise in both the hexatic phase, or instead, in tilted phases of membranes, the stability of which is controlled by thermodynamic parameters like temperature or tension.  A similar class of models has been developed to study the tilt-ordering of ligands which coat the surfaces of solid functionalized nanoparticles of variable shape and size~\cite{nelson2002}.  Finally, recent studies have consider the nematic ordering in dense packing of rods on spherical surfaces, as models of rod-coated droplets~\cite{lowen2016}.

In these systems, continuum descriptions typically consider gradients of an in-plane order parameter, such as the in-plane director ${\bf n}$ defined on the curved 2D mid-surface of the assembly (for example, the mid-plane of thin nematic shell in Fig.~\ref{fig: intrinsiccurv}B).  A common description invokes a 2D generalization of the Frank free energy, $F_{Frust} =\frac{C}{2} \int dA (\partial_i n_j)^2$, where $C$ is an orientational elastic constant (one constant approximation). The effect of orientational frustration becomes more transparent when relating these gradients to an angle field $\theta({\bf x})$ , defined in the tangent planes of the surface,
\begin{equation}
F_{Frust} =\frac{C}{2} \int dA \big|\nabla_\perp \theta + {\bf A} \big|^2,
\end{equation}
where $\nabla_{\perp}$ is an in-plane gradient and ${\bf A}({\bf x})$ is a vector field tangent to the surface, the so-called ``spin connection", that accounts for local rotations of the coordinate frames that define $\theta$ at nearby points~\cite{lubensky_prost1992, nelson}.  This condition satisfies the condition that $\nabla_{\perp} \times {\bf A} = K_g$, representing the build up of orientational gradients on intrinsically-curved surfaces.  For the case of uniform $\theta$, it is straightforward, then, to see that orientational gradients must grow along some in-plane direction, measured by some coordinate, say $y$, as $|{\bf A}| \approx K_g y$, such that the per unit area cost of non-parallelism imposed by geometry grows as $\approx C(K_g y)^2$.

Most models of liquid crystalline order on curved surfaces have considered the effect of topological defects (singular disclinations in the field $\theta({\bf x})$) to ``screen" the orientational elastic effects imposed by curvature.  These studies have explored, specifically, how curved shape or elastic anisotropy control the optimal defect arrangement~\cite{turner2004, shin2008}.  A related problem, also well studied, is the shape relaxation of {\it flexible membranes} due to the presence of disclinations in vesicular assemblies~\cite{lubensky_prost1992}.  Less studied are the the thermodynamic effects of the size dependent elastic costs of orientational gradients.  The arguments above imply an elastic energy density of curved liquid crystalline membranes (in the absence of defects) that grows with lateral dimension as $(K_g w)^2$.  Although this ``orientational elasticity" gives rise to a weaker power law than the quartic scaling of 2D curved crystals, the growth of $F_{Frust}$ with dimensions of is sufficient to give rise to regimes of self-limiting assembly.  The thermodynamic range of self-limiting assembly of liquid crystalline membranes has not yet been studied.

\subsubsection{Metric frustration}

The model introduced in Sec.~\ref{sec: example} illustrates the frustration of positional ordering by Gaussian curvature. Indeed the frustration of positional order is directly related to the frustration of orientational order, since periodic solids have both long-range translational {\it and} orientational order.  The arguments above suggest that skew between straight (geodesic) directions grows along some direction $y$ as $\approx K_g y$.  The convergence or divergence of lattice rows implies straining of inter-row distances that grows as $\approx K_g y^2$, accounting for the power-law difference between costs of orientational and positional strains in assemblies.

The effect of curvature-induced frustration on 2D, positionally ordered assemblies has a long-history, dating back to the Thompson problem~\cite{weaire} which seeks ground states of long-range repulsive particles on spheres~\cite{altschuler1997, perez-garrido1997, cacciuto2002}.  In the context of assemblies, the role of frustration imposed by spherical curvature was recognized by Caspar and Klug in their classification of quasi-icosohedral symmetries of spherical viral capsids~\cite{caspar_klug}.  This and other physical models of capsid assembly~\cite{lidmar2003, zandi2004, nguyen2005, morozov2010} have typically focussed on structure:  on the necessity to incorporate a certain distribution of five-fold coordinated sites (i.e. disclinations) in the protein shell.  Similar considerations have been applied to the formation of clathrin baskets~\cite{mahsl1998, kohyama2003}, 2D mesh-like assemblies of proteins that promote membrane budding through the assembly of positive-curvature scaffolds on the cell wall~\cite{heuser80}.  More recently, the optimal defect pattern in spherical crystals has received attention in the context of particle-coated emulsion droplets~\cite{bausch2003}.  In most such systems, interfacial mediated forces between spherical parts lead to dense, cohesive packings of particles, in predominately 6-fold packing of spheres is punctuated by at least 12 5-fold disclinations.  Experiments~\cite{irvine2010}, simulation~\cite{wales2009, wales2013, burke2015} and theory~\cite{travesset2005, chushak2005, grason_davidovitch2013, azadi2014, azadi2016} on these systems, and others where particles coat hemispherical or even negatively-curved surfaces, reveal complex transitions between compact isolated disclinations, and extended scars of alternating 5- and 7-fold disclinations (see Fig.~\ref{fig: intrinsiccurv}C). 

A smaller body of work has addressed the problem of self-limiting or finite width crystalline domains with intrinsic curvature.  One of the first models to do so was motivated by the appearance of finite width and anisotropic domains of solid phase lipids on (spherical) vesicles~\cite{korlach1999, chen2013}.  In their continuum model of crystalline domains covering partial fractions of spheres, Schneider and Gompper showed that circular domains become unstable to break up into multiple domains of smaller radius as the ratio of line tension to elastic modulus of the crystal drops, and further that circular domains ultimately give way to belt-like ribbon domains above a critical area fraction~\cite{schneider2005}.  Beyond the potential application to solid lipid domains, the 2D crystal continuum model has also been applied to study shape instability of a growing spherical shell, as a model of viral capsid formation~\cite{morozov2010}, where again the isotropic cap becomes unstable to anisotropic shape modes above a critical width.  Recently the shape instability of defect-free crystal domains on spherical surfaces has been observed (see Fig.~\ref{fig: intrinsiccurv}D) in experiments of depletion-mediated assembly of colloids at surface of an oil droplet~\cite{meng2014}.  Experiments observed a transition to branched, ribbon-like domains of finite width that decreased with sphere radius according to the elastic self-limitation mechanism described above.  Motivated by the observations of hierarchical assembly of finite domains, several theoretical studies have been developed to probe the mechanism and detailed structure of the larger-scale structure defect-free crystals on spheres~\cite{gomez2015, voigt2015, voigt2016}.

\subsection{Chirality vs. long-range order in filamentous assemblies}

\begin{figure}
 \epsfig{file=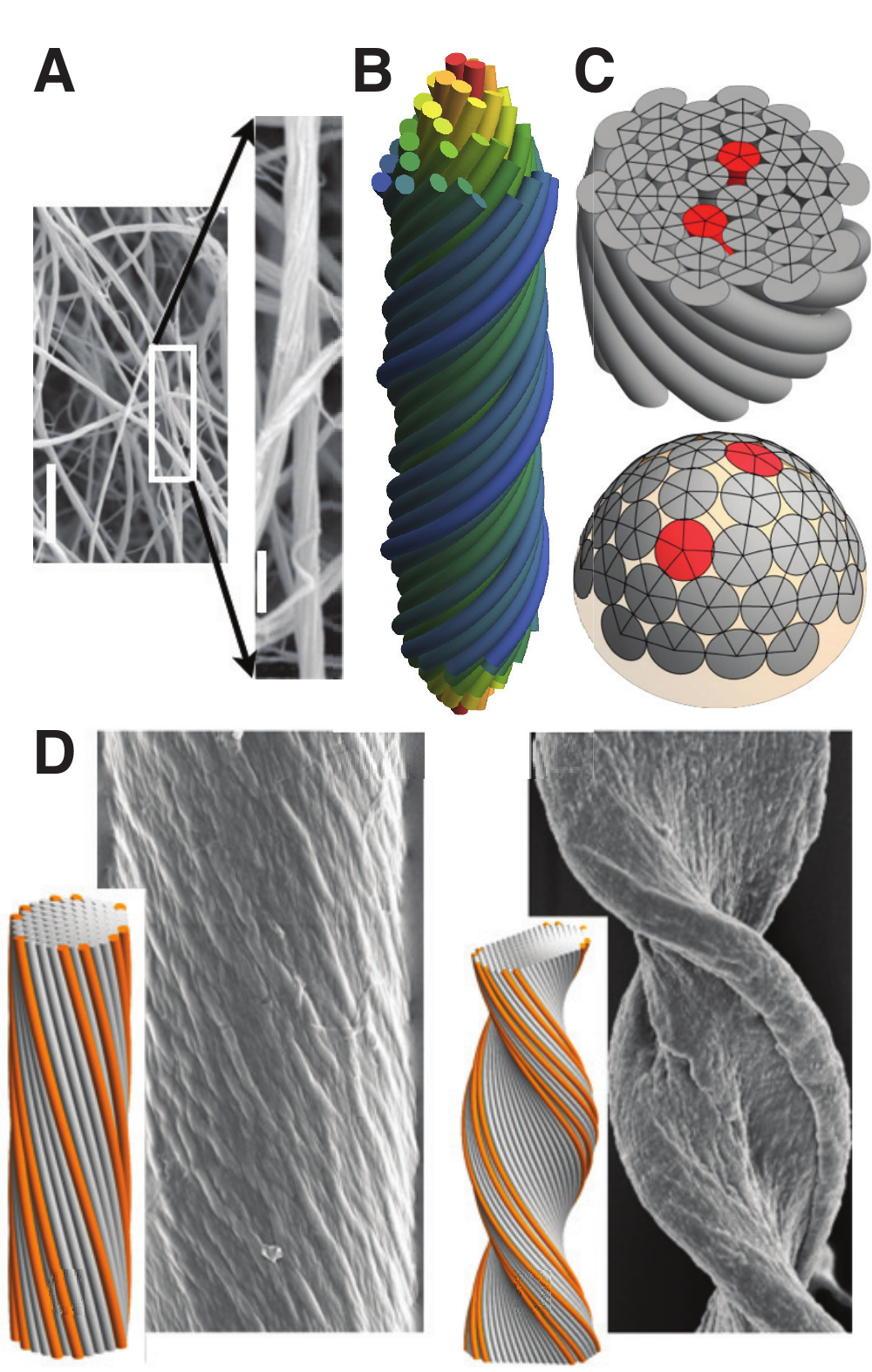, width=3in}%
 \caption{In (A), an electron microscopy image of an artificial blood clot formed by the assembly of fibrin bundles, adapted with permission from J. Thromb. Haemo. {\bf 5}, 116 (2007). Copyright Wiley Publishing.  Scale bar indicates 0.2 ${\rm \mu m}$.  (B) shows a schematic of a ``double-twist" texture of filaments (courtesy of I. Bruss).  (C) shows the results of numerical simulations of twisted bundle packing (top), with 6- and 5- fold filaments shown as grey and red, respectively, and the ``dual surface mapping" of the same packing (bottom), adapted with permission from Soft Matter {\bf 9}, 8327 (2013). Copyright Royal Society of Chemistry (2013).  (C) shows scanning electron micrographs of twisted amyloid showing a morphology distinction between cylindrical (left) and tape-like (right) fibers, adapted with permission from Soft Matter {\bf 8}, 10298 (2012). Copyright Royal Society of Chemistry (2012).  Insets show discrete filaments models of isotropic and anisotropic bundle morphologies (after ref.~\cite{hall2016}). }%
 \label{fig: filaments}
 \end{figure}

The second mode of frustration is broadly defined as the incompatibility between chiral order and long-range orientation/positional ordering in assemblies of chiral filaments.  The transfer of molecular scale chiral structure to the handedness, or screw symmetries, of a self-organized assembly is a broad theme in molecular assembly~\cite{harris1999}.  The ``twisted" textures of rod-like chiral elements are of particular interest in the context of liquid crystals~\cite{straley1976}, where elements are small molecules, and also for biological materials, since helical (chiral) polymers and filamentous proteins are  common (macromolecular) building blocks of inter- and extra-cellular structures in living organisms (e.g Fig.~\ref{fig: filaments}A) from bacteria, to plants and animals~\cite{alberts}.  The self-twisted architecture of chiral  filament assemblies has been a key aspect of structural models of a broad range biological~\cite{kornyshev1997, kornyshev2007, neville, bouligand2008, bouligand1985, ottani2001, ottani2002, charvolin2014, weisel1987, makowski1986, mcdade1993} and synthetic~\cite{che2004, douglas2009, wang2013} assemblies.  Here, I describe models of frustrated chiral assemblies of rods, focussing on long filaments, and again separate the discussion into models that consider chiral frustration of {\it orientational} and {\it positional} ordering.

\subsubsection{Orientational frustration}

When discussing orientational frustration by chirality, it is important to recognize that not all textures induced by chirality frustrate long-range order.  For example, perhaps the most common chiral texture, the cholesteric phase, in which the nematic director rotates helically along a fixed axis in a system, is not frustrated.  In absence of positional ordering cholesteric order forms on arbitrarily large length scales without size-dependent strains.  This is not possible when a chiral system favors a {\it double-twist texture}, which favors handed rotation of the director along two orthogonal directions simultaneously, which is unlike the uniaxial rotation in the cholesteric~\cite{sadoc_mosseri, kleman1989}. This texture occurs along the central axis of a ``double-twist tube", which shows cholesteric-like rotation of the director field along every radial direction that intersects the axis (e.g. Fig.~\ref{fig: filaments}B).  The texture is frustrated in the sense that double-twist is strictly only possible along the central axis, while the director field eventual becomes cholesteric (unaxial) far from the tube core~\cite{kleman1985}.  

The geometric frustration of double-twist textures was first studied in the context of liquid crystal blue phases~\cite{wright_mermin1989}, which appear in a temperature window between the isotropic and cholesteric phases of chiral liquid crystals.  In these bulk phases, the impossibility of filling space uniformly with double-twist~\cite{sethna_wright_mermin1982} leads to defect rich phases in which disclination lines mediate the orientational mismatch between neighboring double-twist tube domains.  More recently, this model has been applied to the formation of finite-size fiber domains of chiral liquid crystalline assemblies, specifically as a model of collagen assembly~\cite{rutenberg2014}.  Nearly the identical approach has been applied to study domain formation in chiral assemblies of cellulose in plant cell walls~\cite{murugesan2015}.  In these models the frustration free energy takes the of the Frank free energy of chiral nematic phase~\cite{degennes_prost},
\begin{multline}
F_{frust} = \frac{1}{2} \int dV \Big\{ K_1 (\nabla \cdot {\bf n} )^2 + K_2 \big[ {\bf n} \cdot ( \nabla \times {\bf n} ) + q_0 \big]^2 \\+ K_3 \big[ ({\bf n} \cdot \nabla) {\bf n} \big]^2 + K_{24}  \nabla \big[  (\nabla \cdot {\bf n} ){\bf n}- ({\bf n} \cdot \nabla) {\bf n} \big] \Big\}
\end{multline}
where ${\bf n}$ is the director field in the fiber (the long axis of constituent ``filaments"), $2 \pi/q_0$ is the preferred pitch of cholesteric order, $K_1$, $K_2$, $K_3$ are the Frank elastic constants for splay, twist and bend, and $K_{24}$ is the saddle-splay constant that, in combination with the chiral term proportional to $q_0$, drives the preference for double-twist in chiral systems.  Considering a single cylindrical domain of infinite length and taking a the director field ${\bf n} = \cos \theta(r) \hat{z} + \sin \theta (r) \hat{\phi}$, it is straightforward to show that the preferred (chiral) twist leads to the linear tilting of the director (filaments) away from the central axis of the domain $\theta(r) \simeq q_0 r/2$.  In this texture, the lines of the director (which are followed by the backbones of filaments) are helical paths of radius $r$ and pitch $4 \pi /q_0$.  Hence, increasing the width of the double-twist domain $R$ requires increased curvature of those lines, leading to a size-dependent bending energy that grows faster than the area, as $\approx K_3 q_0^4 R^4$.  This size-dependent cost can be shown to give rise to self-limiting sizes of chiral liquid crystalline fibers, with diameters that increase with preferred pitch $2 \pi /q_0$.  

This model should be consider an example of ``soft frustration" because the cost to defrustrate the assembly by untwisting is finite.  An infinite width domain is always optimal over the defect-free single domain provided the bulk free energy gain per unit volume exceeds $K_2 q_0^2/2$.  Of course, as with the blue phase, bulk phases may also possess arrays of defects that allow double-twist locally and achieve lower bulk free energy than uniform order.  The transition from self-limiting double-twist domains and bulk (defect-riddled) blue phases in the double-twist assembly model has so far not been addressed.  

\subsubsection{Metric frustration}

Like Gaussian curvature, chiral patterns of twist frustrate positional, as well as orientational, long-range ordering in assemblies.  Specifically, here we review the frustration of {\it columnar ordering} by a preference of cholesteric twist of the column tangents.  Such a model applies to the state of condensed filaments possessing lateral 2D order traverse to the backbone (column), and several variants of this model have been applied to the self-assembly of chiral filament bundles~\cite{turner2003, grason2007, grason2009, yang2010}.  Relative to the polymer nematic model described above, chirality more fundamentally frustrates columnar order.  Uniaxial (e.g. cholesteric) twist of the column tangents requires in-plane shears that grow along the column axis~\cite{kamien_nelson1995, kamien_nelson1996}, and therefore, cholesteric textures introduce an elastic cost that diverges with column length $L$ as $\sim L^3$, which is usually characterized as the divergence of the twist elastic constant when $L \to \infty$~\cite{selinger_bruinsma1991}.  While cholesteric order is prohibitive in columnar phases, double-twist textures are not (i.e. do not introduce length dependent strains), and therefore, chiral interactions between 2D ordered filaments quite naturally favor handed double-twist.

Although double-twist does not require gradients in the inter-column order along the column axis, it does frustrate the packing of filaments {\it transverse} to the columns~\cite{grason2015}.  Intuitively, one can understand this by considering a twisted assembly in which column (filament) positions twist around the central axis of the tubular domain at a rate $\Omega$ (i.e. with a helical pitch $2\pi /\Omega$ along the domain).  As in the fiber geometry describe above, the tilt angle of filaments with respect to the center axis increases with radius as $\theta(r) =\arctan( \Omega r)$.  Radially increasing tilt leads to crowding of nearest neighbors along the azimuthal (tilt) directions~\cite{kamien1996, grason2007}.  For example, assuming that filaments or columns possess uniform separation, $d_0$, in plane normal the central axis (say, in a hexagonal lattice packing), then the 3D spacing (or, distance of closest approach) between azimuthally-separated neighbors is roughly $d_0 \cos \theta(r) < d_0$.  Notably, no  distortions of the packing can fully relax this geometric compression of neighbor distances.  For example, adjusting radial distance of concentric radial ``shells" in order relieve the azimuthal compression requires stretching of neighbor distances along the radial direction.  

Indeed, it can be shown that the gradient texture of filament orientations ${\bf n}({\bf x})$ generated by double-twist imposes strict constraints on inter-filament spacing throughout the cross section that are formally equivalent to those imposed by Gaussian curvature of a 2D surface~\cite{bruss2012, grason2015}.  Specifically, any 2D cross section of a filament assembly maps directly on a 2D curved surface, whose geodesic distances encode the distance of closest approach (or metric) between filament backbones.  The Gaussian curvature of this surface $K_{eff}$ derives directly from the tilt gradients in the cross section~\footnote{Here $\nabla_\perp \times {\bf v}_\perp = \partial_x v_y - \partial_y v_x$, denotes a 2D curl in the plane perpendicular to $\hat{z}$, and likewise $ {\bf v}_\perp \times \nabla_\perp = v_x \partial_y- v_y \partial_x$.},
\begin{equation}
K_{eff} \simeq \frac{1}{2} \nabla_\perp \times \big[ (\nabla_\perp \times {\bf n}_\perp ) {\bf n}_\perp  -  ({\bf n}_\perp  \times \nabla_\perp) {\bf n}_\perp  \big].
\end{equation}
The double twist texture corresponds to positive Gaussian curvature $K_{eff}(\Omega r \ll 1) \simeq + 3 \Omega^2$, consistent with the build-up with inter-filament compression with distance from the center of the bundle.   Due to this geometric connection to 2D surface geometry, the inter-filament pressure $\Pi$ in an elastic theory of 2D-ordered filaments or columns obeys an identical compatibility relation to eq. (\ref{eq: compat}) with the surface curvature replaced by $K_{eff}$.  For an approximately cylindrical twisted bundle of radius $R$, the geometric strain of inter-filament spacing leads to an additional size-dependent cost for bundle formation
\begin{equation}
\frac{F_{frust} }{\pi R^2 L} = 2 K_2 (\Omega - \Omega_0)^2 +\frac{K_3}{4} \Omega^4 R^2 + \frac{Y}{184} (\Omega R)^4 ,
\end{equation}
where $\Omega_0$ is the rate of double-twist preferred by chirality and $Y$ is an elastic modulus for 2D filament array.  Notably,  when bundles exceed a radius proportional to $\sqrt{K_3/Y}$, the cost of inter-filament strain exceeds the cost of bending (intra-filament strain), leading to a penalty for assembly that grows much more rapidly with size (as $\sim R^6$) than for chiral polymer nematic bundles.

\begin{figure*}
 \epsfig{file=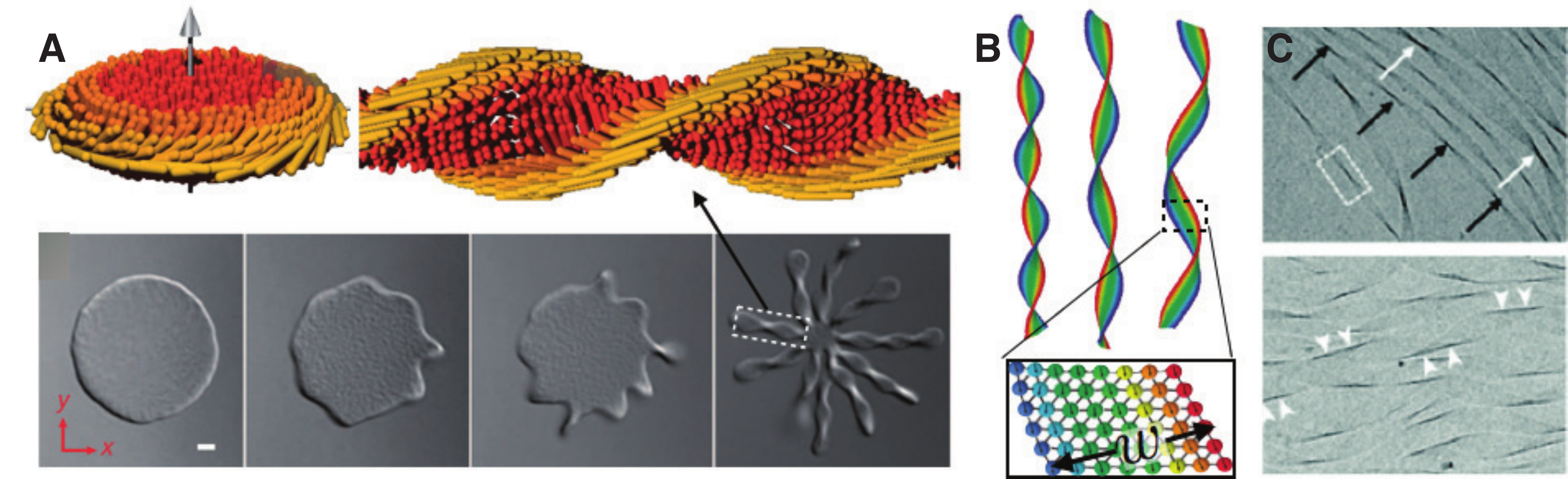, width=6.75in}%
 \caption{In (A), colloidal rafts of condensed fd-virus particles adapted with permission from Nature {\bf 481}, 348 (2011).  Copyright Nature Publishing Group (2011).   From left to right, bottom images show the effect of increasing chirality (decreasing $T$).  At low chirality (top, left schematic) shows twist only at the edge disc-like membrane, while high chirality (top, right schematic) leads to helicoidal twisted ribbons, which emerge spontaneously from an unstable disc.  In (B), a simulation model of chiral crystalline ribbons, showing the effect of increasing ratio of in-plane stiffness to bending modulus from left to right, adapted with permission from Phys. Rev. Lett. {\bf 93}, 158103 (2004). Copyright American Physical Society (2004).  In (C) cryoEM images of chiral lipid bilayers showing helicoidal morphologies for narrow ribbons (top) and isometric spirals for wider ribbons (bottom) (adapted with permission from  J. Am. Chem. Soc. 133, 238105 (2011).  Copyright 2011 American Chemical Society.   }%
 \label{fig: 8}
 \end{figure*}

The rapid radial growth of the inter-filament strain in twisted, 2D ordered bundles leads to smaller self-limited radii with equilibrium radius growing more slowly with bulk chemical potential drive for assembly than the polymer nematic model ($R_{col} \sim (\Delta \mu)^{1/4}$ compared to $R_{LC}\sim (\Delta \mu)^{1/2}$) .   Beyond self-limiting size, the frustration of long-range 2D order in twisted bundles may trigger other structural responses.  Like 2D crystals frustrated by Gaussian curvature, above a critical radius, topological defects become stable in the 2D bulk packing of twisted bundles.  When the tilt angle of outer filaments exceeds a critical value of $\theta(R) \approx 25^\circ$, one or more 5-fold disclinations become favorable in the cross sectional packing~\cite{grason2010, grason2012, bruss2012, bruss2013}.  Excess 5-fold defects, characteristic of {\it positive} Gaussian curvature surfaces, have been observed in multiple numerical simulation models of twisted bundles (see Fig.~\ref{fig: filaments}C), both when twist emerges from chiral inter filament forces~\cite{yang2010, bruss2013} and when twist is imposed mechanically on otherwise achiral bundles~\cite{dumitrica2015}.  Another manifestation of frustration in twisted bundles is the instability of the cross-sectional shape of bundles.  Paralleling the cap-to-ribbon transition of spherical curved crystals, a recent model and simulation show that round bundle cross sections become unstable to anisotropic domain shapes~\cite{hall2016}.  This results in twisted tape-like bundles, when the ratio of inter-filament elastic strain to filament bending energy exceeds a critical value, where the narrow dimension of the tape is limited by inter-filament strain.  Compared to the 2D curved crystal model~\cite{schneider2005}, the anisotropic domains of twisted bundles have the additional costs of filament bending, which leads to the limitation of the larger dimension of the cross section by the cost of filament bending.  This model has been applied to observations of morphological differentiation of amyloid fiber assemblies (see Fig.~\ref{fig: filaments}D) into twisted cylinder vs. helical tape-like bundles~\cite{barone2012, barone2013}.

Models of chiral filament assemblies have been developed to explain size-selection of bundles and fibers of multiple biofilament types, including fibrin~\cite{weisel1987} (e.g. Fig.~\ref{fig: filaments}A) and sickle-hemoglobin microfibers~\cite{turner2003}.  These and other models consider the additional thermodynamic costs imposed by 3D solid order in bundles, which possess additional translational ordering along the filaments or columns~\cite{charvolin2014}.  These models show this additional ``layer" of ordering imposes a free-energy penalty for inter-filament sliding shears that grows as $\Omega^2 R^4$~\cite{grason2007, grason2009}, and further additional cost of the non-linear longitudinal stretching of filaments required to maintain ``smectic-like" ordering in bundles, that grows as $\Omega^4 R^6$~\cite{turner2003}.  Notably this latter model generates the same twist- and size-dependent cost as is already present due to frustration of 2D filament packing, making it challenging to  distinguish between predictions of 2D columnar and 3D solid (stretch-dominated) models for self-limiting radius of cylindrical bundles alone.  The distinctions between the models should be more apparent at weaker twists where sliding shear dominates over both higher-order effects~\cite{grason2009}, or instead, from the appearance of other structural instabilities like defects in the lateral bulk order or anisotropic domain shapes.

\subsection{Chirality vs. layer formation in 2D assemblies}
The third mode of geometric frustration in assemblies is the formation of 2D membranes with chiral order.  The most commonly studied examples are 2D membrane formation of rod-like elements~\cite{langer_sethna1986, selinger_mackintosh_schnur1996, selinger2001}, whose chiral structure promote twisted-nematic textures in the layers.  The frustration of chiral order in 2D layer geometries applies more generally to the case of any 2D assembly possessing in-plane orientational order, whether that in-plane anisotropy derives from tilted-nematic order or not.

The antagonism of chiral ordering and layering was described by deGennes by way of an analogy between smectic liquid crystals and superconductivity~\cite{degennes1972}.  In many layered phases (smectics-A) the nematic director ${\bf n}$ prefers to point along the layer normal ${\bf N}$, but layer normals cannot twist, at least not in the sense of cholesteric order.  In the simplest description, this leads to free energy cost of the form
\begin{equation}
F_{frust} = \frac{1}{2} \int dA \Big\{K_2\big[ {\bf n} \cdot (\nabla \times {\bf n}) +q_0 \big]^2 - C ({\bf N}\cdot {\bf n})^2 \Big\}
\end{equation}
where the second term introduces a generic penalty for nematic tilt away from ${\bf N}$, and we have dropped splay and bend terms for clarity.  Chirality favors $ \nabla \times {\bf n}=-q_0  {\bf n}$, but for any 2D surface, it is straightforward to show that $ \nabla \times {\bf N}=0$, so threading cholesteric twist into a membrane requires directors to break away from the normal.  In bulk chiral smectics, Renn and Lubensky showed that for sufficiently strong chirality, this can occur through the introduction of defects, twist-grain boundary arrays of dislocations~\cite{goodby2002}, which locally disrupt the layered order and allow chiral twist of the director in the defect vicinity~\cite{renn_lubensky1988, lubensky_renn1990}.  In layered assemblies of finite dimension, cholesteric twist can exist at a free edge of any assembly by tipping the director down over some distance $\lambda$ near the layer edge~\cite{pelcovits_meyer2009}.  In this model, this distance is set by the balance between elastic costs of twist and deflections from the normal $\lambda = \sqrt{K_2/C}$ such than maximum tilt angle at the edge is roughly $\theta_m \approx \lambda q_0$.  Since anchoring to the normal prevents cholesteric twist in the bulk, the twist allowed at the free edge lowers the effective line tension by roughly $\Delta \Lambda \approx - K_2 \lambda q_0^2$.  

A manifestation of the the relaxation of frustrated twist at the boundary of membrane was observed in experiments by Dogic and coworkers in condensed, 2D membrane phases of chiral rods, filamentous fd viruses~\cite{gibaud2011}.  In these systems, the effect of chirality, as quantified by the inverse bulk pitch $q_0$, was observed to be temperature dependent.  For weak chirality (i.e. low $|q_0|$), large 2D membranes (see Fig.~\ref{fig: 8}) with rod axes predominantly aligned normal to the membrane, except within a zone near to the free edge permitting twist penetration~\cite{barry2009} .  The circular shape and uncontrolled radius of these flat membranes derives from a positive total line tension.  In accordance with the argument summarized above, at sufficiently high $q_0$, twist penetration at the boundary drives the effective line tension below zero, and destabilizes the perimeter-minimizing shapes of disc membranes.  Negative line tension leads to proliferation of free boundary through the formations of narrow ribbons that split-off in ``starfish" patterns from central disc-like membrane.  Rather than maintaining flat ribbon and twisting along the width of the ribbon, the mid-plane of these ribbons adopts a helicoidal shape that facilities twist along the long axis of the ribbon.  Hence, in the highly-frustrated, strong chirality regime, membranes become ``all edge" to facilitate cholesteric twist throughout the assembly.

A similar mechanism has been implicated in the formation of helical nanoribbons in smectic phases of bent-core liquid crystals.  The biaxial symmetry of these molecules gives rise to smectic layering that spontaneously breaks chiral symmetry, leading to a drive for cholesteric twist in the membrane that parallels the behavior of chiral rod membranes~\cite{hough20091}. Chiral order leads to the formation of stacked helicoidal smectic layers, dubbed  {\it helical nanofilaments} (see Fig.~\ref{fig: layer}A), whose width and thickness (stack height) are finite and in the range of 10s of nm, while the length of nanoribbons is unlimited~\cite{hough20092, zhang2014}. The self-limitation of width can be understood by considering the ``high chirality" limit~\cite{matsumoto2009} where $K_2$ is sufficiently large to lock the director into the preferred cholesteric twist, ${\bf n}(z) = \big(\cos (q_0 z) , \sin (q_0 z),0\big)$, where $z$ is the pitch axis of the helicoid and $\hat{w} =\big(- \sin (q_0 z),  \cos (q_0 z), 0 )$ is the direction along its width.  While it is possible to align the director and the normal along the central pitch axis of helicoid, frustration prevents their global alignment and ${\bf N} - {\bf n} \simeq q_0 x_w \hat{z}$, where $x_w\in [-w/2,+w/2]$ is the position along the width.  Thus, we expect a frustration cost $F_{frust}/wL \approx C (q_0 w)^2$ to compete the line tension drive toward $w \to \infty$, thus setting the finite width of helical nanoribbons.  The mechanism that sets the finite thickness (stack number) is discussed in the follow subsection.

Crystalline order in membrane assembles adds a further layer of frustration for chiral ordering.  This has been studied predominantly in the context of bilayer assemblies with chiral tilt order~\cite{selinger_mackintosh_schnur1996, selinger2001}, often as model of membranes formed by chiral amphiphiles~\cite{oda2003}.  As above, chiral ordering relies on in-plane anisotropy that may be described by an in-plane vector ${\bf c}$, say the in-plane tilt of SmC layer.  Assuming components of ${\bf c}$ are fixed in the plane, Helfrich and Prost~\cite{helfrich1988} showed that the coupling of membrane shape to chirality has the form $c_i  \kappa_{ij} b_j$, where $\kappa_{ij}$ is the curvature tensor and ${\bf b}= {\bf N} \times {\bf c}$, such that, membrane chirality generically drives off-diagonal curvature $\kappa_{cb} \approx q_0$~\cite{selinger2001}.  In chiral crystalline membranes, where $K_g=0$ requires in-plane elastic stretching, the preference for chiral curvature drives a width-dependent shape transition~\cite{selinger2004, ghafouri2005, ziserman2011, arman2014}. In 2D bilayer ribbons where ${\bf c}$ is along or parallel to the long axis, the bending also favors zero mean curvature, or $\kappa_{cc} = - \kappa_{bb}$, which implies ribbon configuration with {\it negative Gaussian curvature} $K_g = - \kappa_{cc}^2 - \kappa_{cb}^2 \approx -q_0^2$.    Thus, when bending energy dominates for sufficiently narrow ribbons, membranes adopt minimal surface, helicoidal shapes, which are accompanied by an elastic penalty due to frustration of 2D positional order $F_{frust}/wL \approx Y (q_0 w)^4$.  This width dependent cost has been predicted to lead to self-limiting ribbons widths, provided that equilibrium width does not grow too large, which occurs for sufficiently small chemical potential drive.   As ribbons grow wider, frustration cost drives the minimal surfaces to transition to {\it isometric} spiral configurations (i.e. $K_g =0; H \neq 0$) at which point $F_{frust}/wL \approx B q_0^2$ ($B$ is the bending stiffness) is no longer width dependent and the $w$ is not self-limiting in equilibrium~\cite{ghafouri2005, arman2014}.  In this way, the transition from helicoidal to spiral ribbon for 2D solid chiral membranes may be viewed as an example of the escape of ``soft frustration", controlled by the relative cost of in-plane stretching to membrane bending $\sim Y q_0^2 w^4/B$.

\subsection{Curvature vs. uniform stacking in multi-layered assemblies }

The fourth mode of geometric frustration derives from the incompatibility of curvature and even spacing in multi-layered assemblies.  The most commonly studied examples of multi-layered assemblies have smectic liquid crystal order:  stacks of 2D fluid layers.  Geometry leads to a non-linear coupling between layer curvature and stack spacing.  The elastic energy of smectic assemblies favors equal spacing between layers, $d$ both along the layer direction and the stack direction.  Consider a smectic layer with mean and Gaussian curvatures, $H$ and $K_g$ to which we add a layer spaced a distance $z$ along the normal direction of surface.  The mean and Gaussian curvature of that layer at $z$ becomes~\cite{didonna_kamien2002, didonna_kamien2003},
\begin{equation}
\label{eq: H(z)}
H(z) = \frac{H+K_g z}{1+2 H z + K_g z^2} ; \ K_g (z) = \frac{K_g }{1+2 H z + K_g z^2}.
\end{equation}
Thus, if the initial layer is curved ($H\neq0$ or $K_g \neq0$), layers spaced at integer multiples of $d$ (i.e. $z=nd$), are also curved, but with {\it different values of curvature}.   This illustrates the fact that it is not possible to simultaneously maintain a constant curved shape (fixed values of $H(z)$ and $K_g(z)$) while at the same time maintaining constant layer spacing.  Equal layer spacing requires layer curvature to evolve, increase or decrease with $z$, and leading ultimately divergence along singular sets at height $z_f$, where $1+2 H z_f+ K_g z_f^2=0$, an effect known as ``curvature focussing"~\cite{sethna_kleman1982}.

\begin{figure}
 \epsfig{file=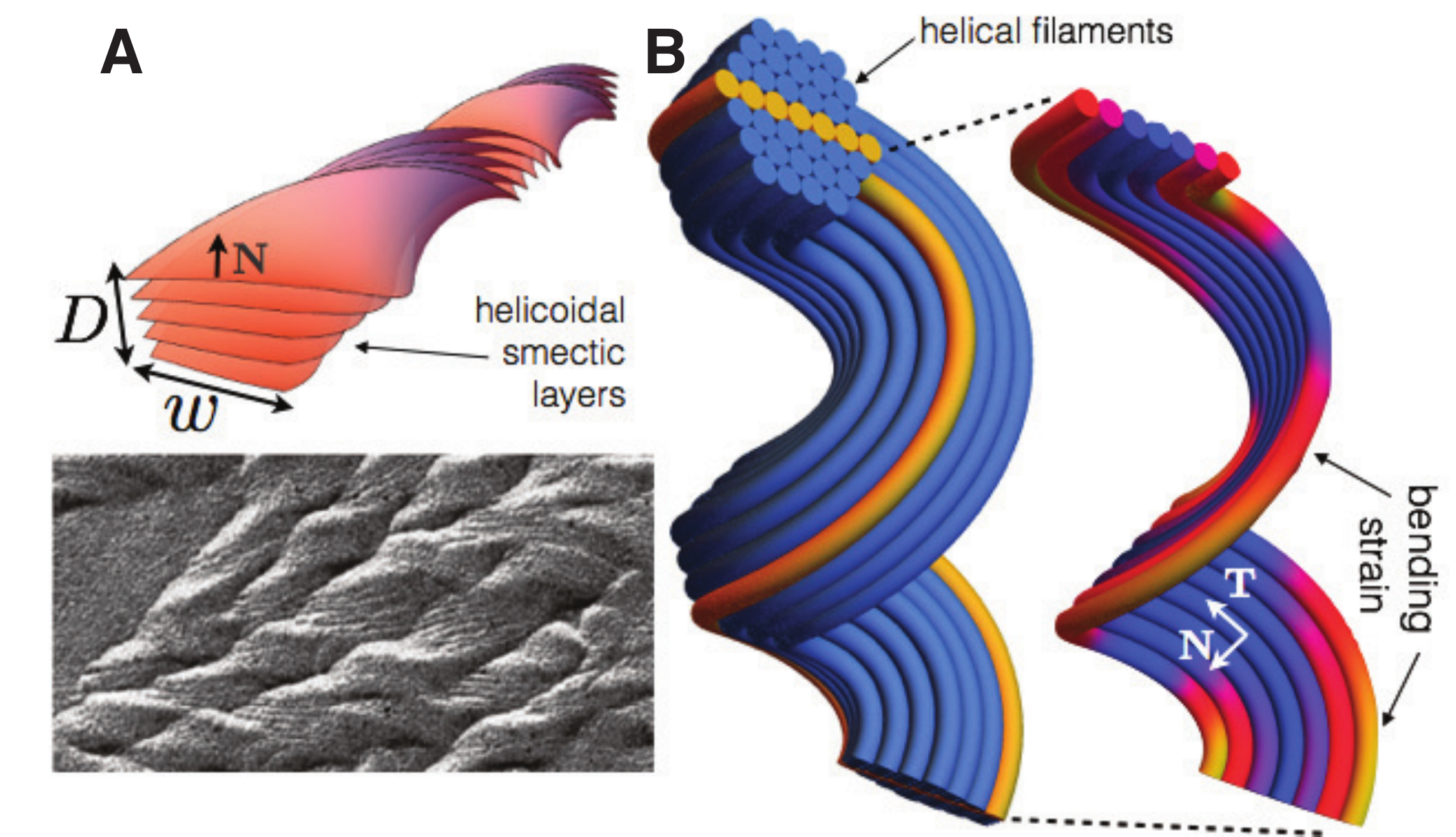, width=3.45in}%
 \caption{In (A), helical nanofilaments of bent core liquid crystals.  The stacking of helicoidal smectic layers in shown schematically (top) and the electron micrograph shows multiple a bulk morphology of multiple nanofilaments, adapted with permission from Langmuir {\bf 26}, 15541 (2010). Copyright 2010 American Chemical Society (2010).  In (B), a schematic of a model of cohesive bundles of helically-shaped filaments, with an extracted row of filaments (right) colored according to bending away from ideal helix shape, adapted with permission from Soft Matter {\bf 9}, 6761 (2013).  Copyright Royal Society of Chemistry (2013).}%
 \label{fig: layer}
 \end{figure}

Several models have explored the consequences of this frustration for smectic assemblies with an intrinsic preference for either non-zero mean or Gaussian curvature, or both.  The simplest models consider an elastic energy of the form,
\begin{equation}
F_{frust} = \frac{1}{2} \int dV ~ \Big[ B(H(z) -H_0)^2 + C (K_g(z) -K_0)^2 \Big],
\end{equation}
where $B$ and $C$ are elastic moduli for deviations from a locally preferred geometry $H= H_0$ and $K_g=K_0$, while the cost of Gaussian curvature is often neglected as a higher-order shape derivative~\footnote{Dimensional considerations suggest that the ratio of elastic moduli $C/B \sim d^2$.  Hence the cost of $K_g \neq K_0$ is generically considered less relevant than mean-curvature changes when curvature radii are much larger than microscopic layer dimension $d$.}. Note also that we neglect, for simplicity, the possibility to adjust inter-layer spacing from $d$, meaning eq. (\ref{eq: H(z)}) applies.  

The role of curvature frustration for the case of cylindrical assemblies ($H_0\neq0; K_0 =0$) has been explored in the context of myelin figures, concentric tubular multilayer stacks that emerge spontaneously from lipid aggregates in drying surface suspensions~\cite{zou2006}.  In this model, self-limitation of a the coaxial stack of thickness $D$ can be understood from the variation of the layer curvature from a $z=0$, $H(z) -H_0 \simeq 2 H_0^2 z$ leading to a strain energy that grows with stack thickness as $H_0^2 D^3$~\cite{santangelo2002, huang2005}.  A second class of applications considers layers with preferred zero mean curvature $H_0=0$, but with non-zero Gaussian curvature $K_0 <0$.   A natural geometry that realizes this shape, at least locally, is the helicoidal tape.  This second class of models has been used to study the self-limiting assembly of helicoidal smectic nanofilaments (Fig.~\ref{fig: layer}A) of bent core liquid crystals~\cite{achard2005, matsumoto2009}, as well as fibrillar stacks of helicoidal tapes formed by amyloid proteins~\cite{aggeli2001}, where interactions between chiral cross-$\beta$ strands of the tape are assumed to drive the handed twist of helical tapes.  For this case $(H_0=0; K_0 <0)$, the mean-curvature strain still grows with stack height since $H(z) \simeq K_0 z$, leading to the same $D^3$ growth of $F_{frust}$ as the $H_0 \neq 0; K_0 =0$ model.  For both models, the competition between size-dependent mechanical costs of curvature change and the cohesive drive to add layers to the stack set a self-limiting thickness for sufficiently weak cohesion.  

The incompatibility of surface curvature and equal spacing in layered systems also has implication beyond curved smectic stacks.  For example, systems with columnar order are also layered systems.  Two-dimensional order transverse to the column axis implies periodicity of density in two independent directions (i.e. the Bragg planes) transverse to the column axis.  Hence, we can decompose a 2D ordered column packing, such as an ordered assembly filaments, in terms of layers composed of 2D columnar rows.  Perfect 2D order implies regular spacing between these layers (rows) as well as between the filaments in the rows.  Due to this underlying layered structure of columnar assemblies, assemblies of curved 1D elements, such as curved filaments, lead to a frustration of 2D columnar order, analogous to the frustration in curved smectics.  The constraints of even spacing in filament assemblies, which we call ``isometric packings"~\cite{starostin2006, grason2009, grason2013} or elsewhere ``developable domains"~\cite{kleman1980, bouligand1980}, lead to conditions on the variation of filament shape throughout a columnar packing.  Locally-curved filament geometry is described by the tangent and normal vectors, ${\bf T}$ and ${\bf N}$, to the backbone (see schematic in Fig.~\ref{fig: layer}B), the curvature $\kappa$ and the torsion, $\tau$, that describes the non-planarity of curves.  In an isometric packing around a given curved filament, the curvature and torsion of neighbor filaments must vary as~\cite{grason2009, grason2013},
\begin{equation}
\kappa ({\bf r}) = \frac{\kappa}{1+\kappa ({\bf r} \cdot {\bf N} )} ; \tau ({\bf r}) = \frac{\tau}{1+\kappa ({\bf r} \cdot {\bf N} )} ,
\end{equation}
where ${\bf r}$ is the distance from a point on reference filament with local curvature $\kappa$ and torsion $\tau$.  Curved filaments that locally separated along their normals must have different curvatures.  While the most commonly studied filament models assume filaments to be intrinsically straight (i.e. with preferred curvature $\kappa_0=0$), there are numerous known examples synthetic or biological filaments with curved backbones in their stress-free configuration.  For example, the analogues of the bacterial cytoskeletal proteins~\cite{ausmees2003, margolin2009}, crescentin, MreB and FtsZ, are believed to assemble into intrinsically curved filaments, while the bacterial flagella is a protein filament that assumes a range of helically-shaped polymorphs~\cite{namba1997}.  Unlike assemblies of intrinsically straight filaments (in the absence of chiral inter-filament forces), even spacing in curved filaments requires shape variation throughout the assembly.  The associated mechanical costs grow with the square of shaped variation from preferred shape, for example, for bending as $\delta \kappa = \kappa({\bf r}) - \kappa_0\simeq \kappa_0^2 r$, where $r$ is the lateral distance from an ideally-shaped filament (see Fig.~\ref{fig: layer}B).  Following these arguments, the elastic energy of isometrically-spaced filaments has been predicted to limit the width to bundles of helically-shaped filaments when the cost of filament shape change relative to inter-filament spacing change is sufficiently high~\cite{grason2009, grason2013}.

\begin{figure}
 \epsfig{file=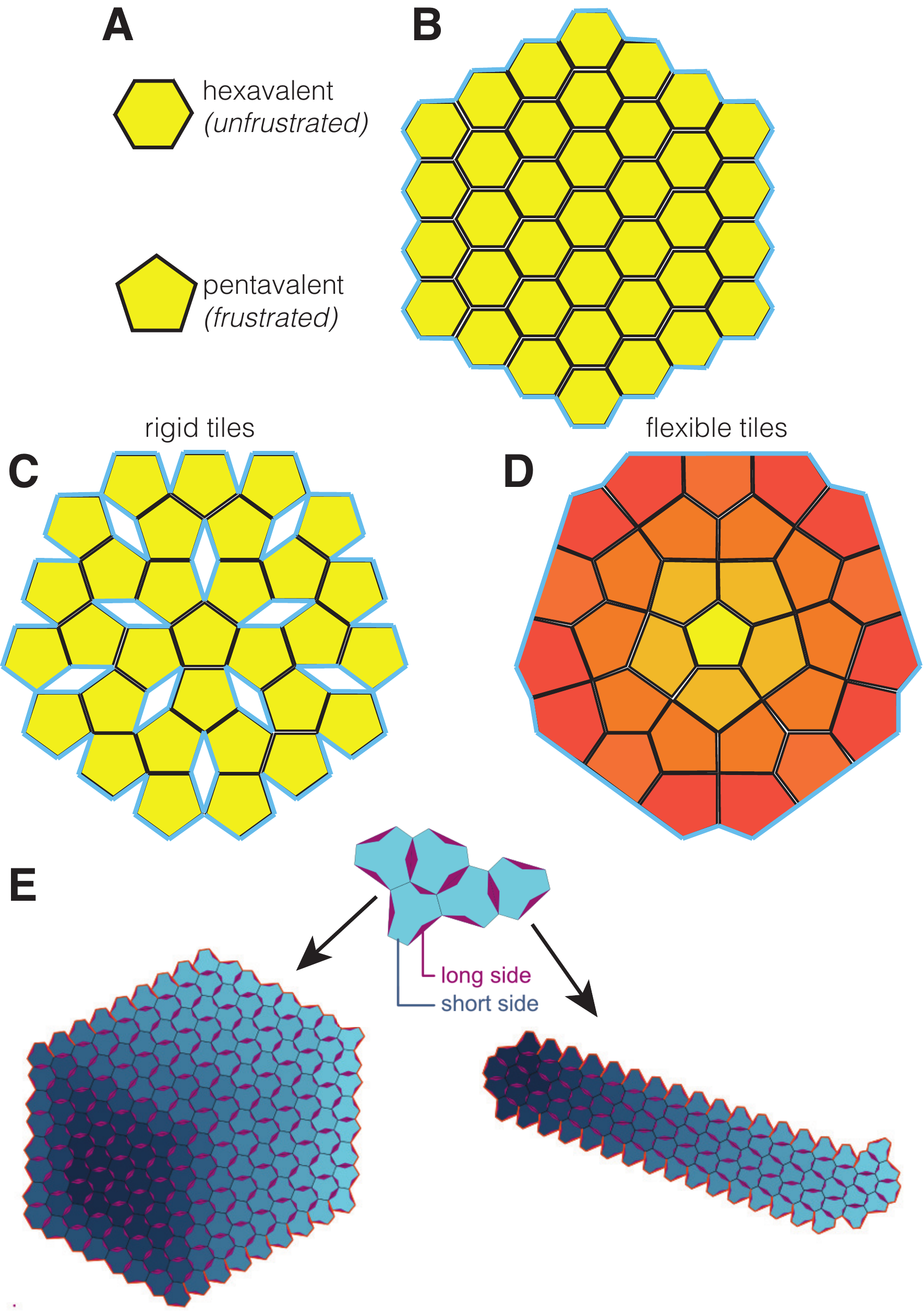, width=2.85in}%
 \caption{In (A) a schematic of hexavalent and pentavalent building block model~\cite{lenz2016} that form cohesive bonds between adjoining faces.  In (B), an example of the unfrustrated assembly of hexavalent units, where rigid elements only lack cohesive bonds at the exterior of an assembled clusters (shown in blue).  In (C), a corresponding rigid cluster assembly of frustrated pentavalent units, which lacks additional cohesive bonds on the cluster interior.  In (D), an illustration of soft pentavalent tile cluster, where tile deformation allows interior bonds to form.  Magnitude of tile strain is depicted with color:  yellow is undeformed and red is highly deformed.  In (E), the model of incompatible, ``pinch hexagons" showing assembly to roughly isotropic clusters (left) for weak line energy and ribbon or ``fiber" assembly for large line energy (right), where shading represents temporal order of the assembly, with ``older" (``newer") blocks shaded dark (light) (adapted from ref.~\cite{lenz2016}, courtesy of M. Lenz).   }%
 \label{fig: zvalent}
 \end{figure}

\subsection{Incompatible shape vs. long-range order in 2D assemblies}

The final mode of geometric frustration describes the frustration of long-range order in assemblies by the shape or valence of interactions self-assembling building blocks.  Compared to the previously introduced modes, the role of this frustration mechanism on self-assembly behavior is far less studied.  Here, frustration derives from the shape of an assembly sub-unit that possesses a symmetry that is incompatible with uniform tiling of the assembly space.  While the many models of assembly consider of locally-isotropic elements spheres or cylinders, many assembly units possess a {\it valence} $z$ of cohesive binding to neighbors.  Examples include ``patchy" particles possessing $z$ symmetrically distributed attractive zones~\cite{glotzer2007, pine2012}, $z$-fold polygonal~\cite{mason2009, mason2011, mason2012} or polyhedral particles~\cite{glotzer2012}, or $z$-fold protein complexes like capsomer assemblies~\cite{zandi2004}, clathrin triskeletal units, or mechano-sensitive membrane channels~\cite{haselwandter2013, kahraman2016}.  

Consider first an assembly of $z$-valent elements confined to the 2D plane that form cohesive bonds with symmetrically distributed neighbors.  Euler's theorem allows for uniform (lattice) tilings only for $z=2,3,4$ and $6$, while other values of $z$, say for penta- or hepta-valent elements, would necessarily introduce gaps or overlap in rigid packings on the plane, and do not admit periodic tilings.  One possible result of the incompatible symmetry of blocks is the formation of  quasicrystalline~\cite{chaikin_lubensky}, or otherwise aperiodic, bulk order perforated by an extensive (i.e. growing with area) distribution of voids, or inter-block gaps.  When $z$-valent elements or their interactions are sufficiently deformable so that the inter-block assembly acquires some elasticity, another behavior is possible, in which frustration is leads to gradient deformations of block shape.  

The assembly of ``elastic", incompatible blocks is considered in a model introduced by Lenz, Efrati and Witten~\cite{lenz2016}.  In one model, they consider the assembly of pentavalent ($z=5$) elastic ``tiles" on the plane (Fig.~\ref{fig: zvalent}A).  In the model, inter-tile cohesion require perfect contact between opposing faces, such that an assembly of rigid pentagonal tiles possessing a distribution of non-bonded tiles interior to the cluster, as shown schematically in Fig.~\ref{fig: zvalent}C.  Alternatively, when tiles are flexible, deformation of the tiles (local shears and dilations) may bridge those non-bonding internal neighbors, leading to a cluster of non-uniform pentagonal tiles like that shown schematically in Fig.~\ref{fig: zvalent}D where non-bonded tiles appear only at the boundary of this cluster.  It may further be anticipated that the elastic cost of shape deformation required by inter-block contact is an increasing function of cluster size.  For example, a crude estimate of the elastic costs would follow by modeling pentavalent block clusters as 2D hexagonal crystals possessing a uniform areal density of 5-fold disclinations.  Balancing the size-dependent cost of tile deformation with the bulk or edge free energy of the cluster, one expects pentavalent tiles to form self-liming clusters, or ribbons (anisotropic domains), provided that blocks are sufficiently deformable relative to the cohesive interaction.  Of course, this scenario neglects the possibility of defects in the 5-valent packing that may tend to localize the cost of frustration and largely mitigate the size-dependent assembly costs.

In their numerical model, Lenz, Efrati and Witten show that anisotropic assembly is possible for a broader class incompatible block.  In particular, they consider hexagonal tiles with alternately long and short edges (i.e. ``pinched hexagons" shown in Fig.~\ref{fig: zvalent}E), such that extending a cohesive tiling over the plane requires block deformation.  In contrast to the angular strains imposed by assembly of pentavalent blocks, defect-free assembly of pinched hexagons will not necessarily introduce block deformations that grow arbitrarily large with size.  Nonetheless, fewer constraints at free edge of a cluster allow some degree of shape relaxation within some zone near the boundary.  Like the case of twist relaxation near the edge of chiral membrane, this has the effect of lowering the effective line energy of the boundary, and potentially promoting the assembly ``all boundary" tape, or fiber-like, morphologies (Fig.~\ref{fig: zvalent}E).  

Among the many new questions opened up by this model are, how the does the geometry of tile shape control the ``strength" and scale dependence inter-block frustration and what are the polyhedral analogs of incompatible tiles that potentially give rise to self-limiting and/or anisotropic assembly regimes in 3D?

\section{Open challenges}
\label{sec: open}

The forgoing sections attempt a classification of the distinct mechanisms of GFA studied to date and survey of the diverse systems in which frustration shapes assembly properties.  This objective points to an obvious open question, what additional modes of frustration can be added to this classification, and what are the self-assembly systems that might realize these modes?  And while this classification of GFA is surely incomplete, additional challenges remain to understanding and manipulating the consequences of GFA in known systems.  Here, I outline a few of those challenges.  

\subsection{Competing structural responses to frustration}
Sections \ref{sec: example} - \ref{sec: modes} outline how frustration of uniform order may lead to distinct structural responses, which reshape the equilibrium size, the boundary shape or bulk order of assemblies.  While the basic principles underlying each distinct response  are fairly well established, it is far from understood how these distinct responses compete in the broader assembly phase diagrams of frustrated systems.  For example, in 2D ordered chiral filament bundles, the boundary between isotropic cylindrical and anisotropic tape-like bundles in defect-free bundles can be related the the relative cost of inter-filament strain to intra-filament bending the bundles, a ratio that grows with surface energy~\cite{hall2016}.  However, it is also known that above a critical twist angle cylindrical bundles become unstable the formation topological defects (disclinations and dislocations) in the interior of the bundle, which lower the overall inter-filament elastic cost~\cite{grason2015}.  Therefore, while the transition from cylinders to tapes in low-twist bundles may take place in the absence of defects, this is surely not the case for high-twist bundles.  In this regime, where both responses are possible it is necessary to determine which response (defects or anisotropic shape), or combination thereof, best mitigates the cost of frustration.  To date, the phase boundaries between isotropic and anisotropic domains in the defect-rich regions of GFAs, such as twisted bundles and curved 2D crystals, remain unknown.  Furthermore, present models do no yet consider the possibility of stable bulk defects altering the equilibrium perimeter shapes to even lower symmetry structures.  

\subsection{Energy landscapes and non-equilibrium behaviors}
Charting the structure of the complex energy landscape of GFAs remains an open challenge. Frustration in condensed matter systems typically implies the existence of multiple, often nearly degenerate, local minima, potentially separated by large energy barriers~\cite{wales}.  In the case of  assemblies, the multiple competing structural responses to frustration pose a particular set a questions about the hierarchical structure of the of the energy landscape.  What are the conditions for metastability of structurally distinct states (e.g. defects vs. anisotropic domains)?  And, when two structurally-distinct minima do coexist for some conditions, are competing minima clustered in regions of the configuration space, or within characteristic energy ``hyper-basins", according the distinct structural response?  As an example of the latter, Wales and Kusumaatmaja have studied the energy minimal of particles and negatively curved interfaces~\cite{wales2013}, whose ground states may possess both ``charged" (with excess disclinations) and ``neutral" (with only 5-7 disclination pairs) defect patterns.  By analyzing the disconnectivity graph associated with transition states between minima, they find that the energy landscape largely separates into two large branches (or hyper basins), charged and neutral, implying that a fairly large barrier is required to climb from class of structural minima to the other.  

Understanding the structure of the energy landscape of GFAs has important implications for non-equilibrium assembly properties.  In many, if not most, physical realizations of GFAs, assembly takes place out of equilibrium, and in some cases far from it.  To date, the anomalous behaviors predicted from GFAs are derived from their equilibrium structures, which are shown to be self-limiting and heterogeneous.  It remains to be understood how these same systems behave when driven out of equilibrium, say via a rapid quench from the disassembled to the condensed state.  For example, can systems predicted to be self-limiting in equilibrium be kinetically driven to undergo bulk, unlimited assemblies?  Or instead, will  systems that should escape ``soft" frustration in equilibrium funneled towards self-limiting energy minima by the global structure of the energy landscape?  Finally, the role of kinetics in GFAs while so far unexplored, should have significant consequences for structure formation in these systems.  As equilibrium states of GFAs adjust their structure as each self-assembling unit is added, even the simplest kinetic model must compare the rate of subunit addition to that of one or more structural relaxations of the assembly.   Models have not yet explored which features distinguish the structures during ``overdamped" assembly from the equilibrium states of GFAs.

\subsection{Engineering geometrically-frustrated assembly by design}
Beyond the foregoing open challenges to our understanding of GFA behavior, a major challenge remains to develop geometric-frustration as a ``rational design principle" for synthetic self-assembling materials.  In current realizations of GFA, both synthetic and those uncovered from biology, the principles of frustrated assembly have been applied as {\it a posteri} explanations of complex assembly behavior.  Recent decades have seen an explosion in techniques for controlling the shape and interactions of self-assembling building blocks, from macromolecules and nanoparticles to colloids~\cite{glotzer2007}.  It remains an open challenge to apply these techniques to develop new classes of self-assembling building blocks which are designed intentionally frustrate homogeneous packing, and thereby harness size-dependent strains of GFAs to direct the size and shape of equilibrium assembly.  In biology, the assembly of finite structures from identical sub-units is ubiquitous and functionally vital, from photonic nanostructures, to viral capsids, to size-regulated intra- and extra-cellular fibers.  In synthetic materials, self-limiting (equilibrium) assemblies are restricted to either molecular amphiphiles (e.g. surfactants, copolymers), whose assembly dimensions are limited to molecular dimensions, or instead to ``programmable assemblies" of selectively interacting building blocks (e.g. DNA origami)~\cite{rothemund2006, dietz2009} , which require the number of interacting species to grow with assembly dimensions~\cite{zeravcic2014}.  GFAs, by contrast, would offer unique platforms to direct the self-limiting dimensions of assemblies to be arbitrarily much larger than the building block size, yet requiring only a {\it single building block} species.  

Critical to the challenge of engineering geometric frustration in order to direct the large-scale structure of assemblies {\it a priori}, is the need to control the ``degree" of frustration through shape control, but also the ratio of inter-block cohesion to elastic stiffness of the assembly itself, which selects the ultimate self-limiting size.  While the range of physical interactions between building blocks assembling in aqueous media, say, is limited to the scales of a few nm at best, maximizing the elastic compliance of an assembly to extend the size range of self-limiting assembly, will likely require the design of ``soft" building blocks whose shape deformations carry the elastic load of inter-block misfit.  This will pose further challenges to designing self-assembling objects whose shapes are both sufficiently well-defined to control inter-block frustration, yet compliant enough to tolerate large variations of inter-block contact.  It remains to be determined what synthetic platforms will be best suited for exploring and manipulating the interplay between shape misfit, cohesion and deformability in GFAs.

\begin{acknowledgments}
I am grateful to C. Santangelo, I. Bruss, D. Hall and I. Prasad for helpful discussions and comments on this article.   I am also grateful to M. Lenz and T. Witten for sharing and discussing their unpublished work described here, and also to I. Bruss and V. Manoharan for sharing unpublished figures for use in this article.  This work was supported by NSF CAREER Award DMR 09-55760.  I would like to acknowledge the hospitality of the Kavli Institute for Theoretical Physics (supported under NSF Grant No. PHY11-25915) where this manuscript was completed in part.  This article is dedicated to the memory of W. S. Klug.
\end{acknowledgments}

\bibliography{bib_v2.bib}

\end{document}